\documentclass[11pt, aps, preprint, nofootinbib,superscriptaddress,eqsecnum,titlepage]{revtex4-2}
\usepackage[active]{srcltx}
\usepackage[utf8]{inputenc}
\usepackage{enumerate}
\usepackage{bbold}
\usepackage{manfnt}
\usepackage{physics}
\usepackage{latexsym}
\usepackage{hyperref}
\usepackage{amsmath}
\usepackage{amsfonts}
\usepackage{graphicx}
\usepackage{slashed}
\usepackage{xcolor}
\usepackage{soul}
\usepackage{slashed}
\usepackage{braket}
\usepackage{natbib}
\usepackage{cancel}
\usepackage{MnSymbol}
\usepackage{wasysym}
\usepackage[export]{adjustbox}
\usepackage{graphicx}
\usepackage{caption}
\usepackage{subcaption}

\begin{document}
	
\title{\Large{{\sc An Introduction to Higher-Form Symmetries}}}

\author{Pedro R. S. Gomes}
\email{pedrogomes@uel.br}
\affiliation{Departamento de Física, Universidade Estadual de Londrina,  86057-970, Londrina, PR, Brasil}

	
\begin{abstract}

These notes are intended to be a pedagogical introduction to higher-form symmetries, which are symmetries whose charged objects are extended operators supported on lines, surfaces, and etc. This subject has been one of the most popular and effervescent topics of theoretical physics in recent years. Gauge theories are central in the study of higher-form symmetries, with Wilson and 't Hooft operators corresponding to the charged objects. Along these notes, we discuss in detail some basic aspects, including Abelian Maxwell and Chern-Simons theories, and $SU(N)$ non-Abelian gauge theories. We also discuss spontaneous breaking of higher-form symmetries.

\end{abstract}
	
\maketitle

\tableofcontents
	
	
\section{Introduction}

Recent years have witnessed incredible progress with the discovery of new forms of symmetries, usually referred to as {\it generalized symmetries}. Although the embryo of the generalized symmetries is already contained in some previous works as in \cite{Alford:1991vr,Alford:1990fc,Alford:1992yx,Bucher:1991bc,Pantev:2005rh,Pantev:2005wj,Pantev:2005zs,Hellerman:2006zs}, and also in the splendid Ref. \cite{Nussinov_2009}, by Nussinov $\&$ Ortiz, the foundational paper recognizing all their glory is \cite{Gaiotto:2014kfa}, by Gaiotto, Kapustin, Seiberg, $\&$ Willett, providing a new status to this subject.  

Intensive exploration of these new types of symmetries has led to a very deep and powerful framework, enriched with ideas from different sides of physics like quantum computing, topological phases of matter, quantum field theory, strings, and quantum gravity. In addition to providing several new ideas and insights on a variety of physical systems, the study of new forms symmetries has enforced us to rethink about certain pillars of modern physics. A remarkable example is the UV/IR mixing inherent to the so-called {\it subsystem} symmetries, which is a feature defying one of the central organizational principles of physics, namely, that physics is organized by scales and different scales are decoupled. Other popular types of generalized symmetries are {\it higher-form} and {\it non-invertible} symmetries. Recent overviews that provide a valuable guide to the literature are \cite{McGreevy:2022oyu,Cordova:2022ruw}. A mathematically-oriented review can be found in  \cite{Sharpe:2015mja}. Since the first version of these notes other reviews have appeared in the literature \cite{Schafer-Nameki:2023jdn,Brennan:2023mmt}.

Higher-form symmetries are quite common and are present in many ordinary relativistic theories, for example, in Abelian and non-Abelian gauge theories, and follow from the existence of  completely anti-symmetric conserved currents $J^{[\mu\nu\ldots]}$. The anti-symmetric nature of the indices implies that we can construct conserved charges by integrating over certain subdimensional spatial manifolds, rather than in the whole space as in the case of an ordinary symmetry. As a consequence, the charged objects are no longer local operators, but are instead extended objects (line, surface, and etc). In addition, the charges are topological in the sense that they are independent of coordinates. As we shall discuss extensively along these notes, the topological meaning becomes more transparent by rephrasing the conservation laws in terms of links between geometric objects\footnote{For example, in $D=4$, a line has a nontrivial link with a two-dimensional sphere $S^2$.}.  

Subsystem symmetries are intrinsically connected with the subject of {\it fractons} \cite{Chamon_2005,Haah_2011,Vijay_2016}, which are excitations with restricted mobility appearing in certain exotic phases of matter (recent reviews can be found in \cite{Nandkishore:2018sel,Pretko:2020cko}). They are similar to the higher-form symmetries in that they also lead to conserved charges along certain subdimensional manifolds, but with the crucial difference that the charges are coordinate-dependent. In a discretized (lattice) system, this implies that there are as many charges as the size of the system. This enormous amount of conserved charges leads to a huge degeneracy of the states. In particular, the ground state degeneracy, which is of course a low-energy quantity, is affected by the number of sites that constitute the system. This is an example of the UV/IR mixing mentioned above.  

The notion of a non-invertible symmetry lies in the existence of topological operators which are non-unitary, and consequently do not have an inverse. This type of operator appears abundantly in two-dimensional conformal field theories (CFT) in the form of fusion rules. One of the simplest examples of this structure is the Ising CFT  \cite{Frohlich:2004ef,Frohlich:2009gb}, which contains a non-invertible line $\mathcal{N}$ whose fusion is $\mathcal{N}\times \mathcal{N}=1+\eta$, where $\eta$ is the $\mathbb{Z}_2$ symmetry line. Constraints on the renormalization group (RG) flow can be obtained from the existence of such non-invertible symmetry \cite{Chang:2018iay}. More recently, the existence of non-invertible symmetries has been discovered in four dimensional gauge theories like QED and QCD, where the chiral anomaly operator can be attached to a fractional quantum Hall phase, turning it into a non-invertible operator \cite{Choi:2022jqy}. This construction leads to nontrivial selection rules in the theory. Non-invertible symmetries in the full Standard Model are discussed in \cite{Wang:2021ayd,Putrov:2023jqi}.
 
Generalized symmetries also have important consequences, leading in general to powerful constraints on the dynamics. As in the case of ordinary symmetries, they can be spontaneously broken \cite{Lake:2018dqm,Hofman:2018lfz}, resulting in Goldstone excitations when the generalized symmetry is continuous. They can also have anomalies, in particular, 't Hooft anomalies, which have implications on the IR structure of the theory because of the anomaly matching \cite{tHooft:1979rat}. This enables to uncover valuable information even in the strongly coupled regime, as  in the case of $SU(N)$ gauge theory at $\theta=\pi$, which has a mixed 't Hooft anomaly between time-reversal and the 1-form center symmetry \cite{Gaiotto:2017yup} (see \cite{Wan:2019oyr} for a rigorous derivation of the anomaly from the point of view of a five-dimensional invertible topological field theory).

These notes are entirely dedicated to the higher-form symmetries, and are intended to be a pedagogical introduction to the subject. They are far from being a comprehensive account, but instead focus specifically on certain basic aspects, providing a reasonably detailed exposition. They are organized as follows. In Secs. \ref{Sec2} and \ref{Sec3}, we discuss briefly some aspects of ordinary symmetries that are useful for the later sections. Sec. \ref{Sec4} presents a general introduction to the higher-form symmetries. Secs. \ref{Sec5}, \ref{Sec6}, and \ref{Sec7} are dedicated to the study of higher-form symmetries in Maxwell theory in various dimensions. In Sec. \ref{Sec8}, we discuss the 1-form $\mathbb{Z}_k$ symmetry in $U(1)_k$ Chern-Simons theory. In Sec. \ref{Sec9}, we discuss the 1-form center symmetry in $SU(N)$ gauge theories. Sec. \ref{Sec10} studies spontaneous symmetry breaking of the 1-form symmetry in Maxwell theory. We conclude in Sec. \ref{Sec11} with some final comments. Two appendices summarize useful properties of differential forms and Lie algebras.

		
\section{Aspects of Ordinary Symmetries}\label{Sec2}

In this section we review some basic aspects of ordinary symmetries in quantum field theory, which will be useful in the generalization to the case of higher-form symmetries. This discussion is quite standard and can be found essentially in every QFT textbook, so that we will be brief.
			

\subsection{Symmetries in Classical Field Theory}		

Noether theorem deeply connects continuous symmetries to conservation laws. This relationship can be derived in a simple way. Consider an action $S=\int d^Dx \mathcal{L}(\phi)$ involving a generic set of fields $\phi$'s and assume that under an infinitesimal transformation of the fields, 
\begin{equation}
\phi\rightarrow \phi+\epsilon_a\delta\phi_a,
\label{0.1}
\end{equation} 
with $\epsilon_a$ being a set of constant (global) parameters\footnote{The index $a$ represents generically a set of indices which can be spacetime or internal.}, the action is invariant. This corresponds to a symmetry in the classical theory. 

To find the associated conserved current we promote the global parameters $\epsilon_a$ to local ones, $\epsilon_a\rightarrow\epsilon_a(x)$. In this case, the transformation 
\begin{equation}
\phi\rightarrow \phi+\epsilon_a(x)\delta\phi_a
\label{0.1a}
\end{equation}
is no longer a symmetry. The variation of the action must involve the derivative of the parameters, $\partial_{\mu}\epsilon_a(x)$, which recovers the invariance under global transformation in the case of constant parameters.  Then we can write
\begin{equation}
\delta S=\int d^Dx J_a^{\mu}\partial_{\mu}\epsilon_a(x),
\label{0.2}
\end{equation}
with arbitrary coefficients $J_a^{\mu}$. It is interesting to note how the index structure of the parameters $\epsilon_a$ is reflected in such coefficients (currents). 
Now, the local transformation $\phi\rightarrow \phi+\epsilon_a(x)\delta\phi_a$ can be viewed simply as arbitrary variations of the fields, in which case \eqref{0.2} vanishes upon using the equations of motion, i.e., 
\begin{equation}
\delta S=\int d^Dx J_a^{\mu}\partial_{\mu}\epsilon_a(x)=0~~~(\text{eq. of motion}).
\label{0.3}
\end{equation}
With an integration by parts and using the fact that $\epsilon_a(x)$ are arbitrary, we conclude that the coefficients $J_a^{\mu}$ are actually conserved currents, 
\begin{equation}
\partial_{\mu}J_a^{\mu}=0 ~~~\Rightarrow~~~\frac{d}{dt}Q_a=0,
\label{0.4}
\end{equation}
where we have defined the Noether charges
\begin{equation}
Q_a\equiv \int d^{D-1}x J_a^0 ,
\label{0.5}
\end{equation}
which remain constant along time evolution. 

 Note that the currents $J_a^{\mu}$ are not uniquely determined, as we can always define a new current
\begin{equation}
\tilde{J}_{a}^{\mu}=J_a^{\mu}+\partial_{\nu}\Omega_a^{\mu\nu},
\label{0.5a}
\end{equation}
where $\Omega_a^{\mu\nu}=-\Omega_a^{\nu\mu}$, which are conserved and lead to the same Noether charges \eqref{0.5}.

In the canonical formalism, the Noether charges \eqref{0.5} are the generators of infinitesimal transformations in the sense that their Poisson brackets with some field furnishes the transformation of the field, 
\begin{equation}
\delta \phi_a = \{\phi, Q_a\}.
\label{0.6}
\end{equation}  
The quantum counterpart of this relation replaces the Poisson brackets by commutators. We shall discuss this point in a moment.

\subsection{Symmetries in Quantum Field Theory}

\subsubsection{Canonical Formalism}

The celebrated Wigner theorem (see for example \cite{Weinberg:1995mt}) asserts that in quantum theory the symmetries (not spontaneously broken) must be implemented through either unitary or anti-unitary operators. Unitary operators accommodate both continuous and discrete symmetries, whereas anti-unitary operators serve only for the discrete ones. 

For continuous symmetries, the unitary operator implementing the corresponding transformation can be systematically constructed from the Noether charge:
\begin{equation}
U=e^{i \epsilon_a Q_a}.
\label{0.7}
\end{equation}
They act on the fields as
\begin{equation}
\phi~\rightarrow ~	\phi' = U\, \phi \, U^{\dagger}.
\label{0.8}
\end{equation}
For an infinitesimal transformation, this expression leads to
\begin{equation}
\delta_a \phi = i [Q_a,\phi], 
\label{0.9}
\end{equation}
which is the quantum counterpart of \eqref{0.6}.

For discrete symmetries, even though a Noether charge does not exist, the unitary operator $U$ is meaningful.  For example, consider a real scalar free theory
\begin{equation}
S=\int d^D x \left(\frac12\partial_{\mu}\phi\partial^{\mu}\phi-\frac{m^2}{2}\phi^2 \right).
\label{0.10}
\end{equation}		
This action has a discrete $\mathbb{Z}_2$ symmetry, which acts on the field as $\phi \rightarrow -\phi$. It is implemented through the unitary operator $U$ according to
\begin{equation}
\phi' = U\, \phi \, U^{\dagger} = -\phi.
\label{0.11}
\end{equation}
An explicit form for $U$ can be written in terms of creation and annihilation operators.


\subsubsection{Path Integral and Ward Identities}

The quantum counterpart of the classical conservation laws  \eqref{0.4} are the so-called Ward identities, which lead to relations among correlation functions of the theory. They can be derived in a simple way using path integral. Consider the partition function 
\begin{equation}
Z=\int \mathcal{D}\phi \,e^{iS}. 
\label{1.1}
\end{equation}
Correlation functions are expressed as
\begin{equation}
\langle X\rangle \equiv \frac{1}{Z} \int \mathcal{D}\phi \, X\, e^{iS}, 
\label{1.2}
\end{equation}
where $X$ represents a generic product of fields,  $ X\equiv \prod_j\phi(x_j)$. As the fields are merely integration variables, we are free to rename them or to make changes of the integration variables. First we rename the fields $\phi\rightarrow\phi'$ and then perform a variable changing in the path integral according to \eqref{0.1a}. This amounts to
\begin{eqnarray}
\langle X\rangle &=& \frac{1}{Z} \int \mathcal{D}\phi \phi(x_1)\ldots \phi(x_N)  e^{iS[\phi]}\nonumber\\
 &=& \frac{1}{Z} \int \mathcal{D}\phi' \phi'(x_1)\ldots \phi'(x_N)  e^{iS[\phi']}\nonumber\\
&=& \frac{1}{Z} \int \mathcal{D}\phi \,J  \left(X +\sum_j \phi(x_1)\ldots \epsilon_a \delta\phi_a(x_j)\ldots\phi(x_N) \right) e^{iS+i\delta S}\nonumber\\
&=& \frac{1}{Z} \int \mathcal{D}\phi (1+i\delta A) \left( X +\sum_j \phi(x_1)\ldots \epsilon_a \delta\phi_a(x_j)\ldots\phi(x_N) \right) (1+ i\delta S)  e^{iS} \nonumber\\
&=&  \frac{1}{Z} \int \mathcal{D}\phi \left(X + \sum_j \phi(x_1)\ldots \epsilon_a \delta\phi_a(x_j)\ldots\phi(x_N)+ i X \delta S  + i X \delta A +\cdots \right).
\label{1.3}
\end{eqnarray}
We have admitted a possible Jacobian $J\equiv1+i\delta A$, accounting for eventual anomalies\footnote{See \cite{Arouca:2022psl} for a modern perspective on anomalies.}. Using the variation of the action in the form \eqref{0.2}, we obtain
\begin{eqnarray}
0&=&\int d^Dx \epsilon_a(x) \left[\sum_j   \delta^{(D)}(x-x_j)  \langle  \phi(x_1)\ldots \delta\phi_a(x_j)\ldots\phi(x_N)\rangle \right.\nonumber\\ 
&-& \left. i\partial_{\mu} \langle J^{\mu}_a(x) X \rangle +i \langle \mathcal{O}_a(x) X \rangle\right],
\label{1.4}
\end{eqnarray}	
where we have parametrized the anomaly as $\delta A\equiv \int d^Dx \epsilon_a\mathcal{O}_a$. In the absence of anomalies ($\mathcal{O}_a=0$), the above relation enables us to find the  Ward identities
\begin{equation}
\partial_{\mu}\langle J^{\mu}_a(x) X \rangle= -i \sum_j   \delta^{(D)}(x-x_j)  \langle  \phi(x_1)\ldots \delta\phi_a(x_j)\ldots\phi(x_N)\rangle,
\label{1.5}
\end{equation}		
which provide a set of relations among correlation functions and conservation laws at noncoincident points $(x\neq x_j, \forall j)$. In particular, by integrating both sides over spacetime, we obtain
\begin{equation}
\delta \langle \phi(x_1)\ldots \phi(x_N)\rangle=0,
\label{1.5a}
\end{equation}
which is the reflection of symmetry on the correlation functions. This is a nontrivial statement about symmetry because it is computed as 
\begin{equation}
\int \mathcal{D}\phi\,\delta \left[  \phi(x_1)\ldots \phi(x_N)\right]e^{iS}=0.
\label{1.5b}
\end{equation}
		
The path integral version of \eqref{0.9} can be derived as it follows. Taking $X$ as a single field $\phi(y)$, the relation \eqref{1.5} with $\mathcal{O}_a=0$ reduces to	
\begin{equation}
\partial_{\mu}\langle J^{\mu}_a(x) \phi (y) \rangle = -i \delta^{(D)} (x-y)\langle \delta_a\phi(y) \rangle. 
\label{1.6}
\end{equation}
By integrating $x$ over the region $\mathcal{V}\equiv [y^0-\epsilon,y^0+\epsilon]\times \mathbf{R}^{D-1}$, we get  		
\begin{equation}
\langle Q_a(y^0+\epsilon) \phi(y)\rangle - \langle \phi(y)Q_a(y^0-\epsilon)\rangle=-i \langle \delta_a\phi(y) \rangle, 
\label{1.7}
\end{equation}		
because of the time-ordering inherent to the path integral. In the limit of $\epsilon\rightarrow 0$ the left hand side is identified with the equal-time commutator
\begin{equation}
\langle [Q_a,\phi(y)]\rangle = -i \langle \delta_a\phi(y) \rangle,
\label{1.8}
\end{equation}
which is the expectation value of \eqref{0.9}.


\subsection{Spontaneous Symmetry Breaking and Goldstone Excitations}\label{ssb0}		
		
The Goldstone theorem states that when a global continuous symmetry is spontaneously broken, then there are massless excitations (Goldstone bosons) in the spectrum. This result can be derived directly from the Ward identity \eqref{1.6}. Taking the Fourier transform with respect to $x$, it follows that
\begin{eqnarray}
\int d^Dx e^{ipx}\partial_{\mu}\langle J^{\mu}_a(x) \phi (y) \rangle &=& -i \int d^Dx e^{ipx} \delta^{(D)} (x-y)\langle \delta_a\phi(y) \rangle\nonumber\\
-i \int d^Dx e^{ipx}p_{\mu}\langle J^{\mu}_a(x) \phi (y) \rangle&=& -i e^{ipy}  \langle \delta_a\phi(y) \rangle\nonumber\\
p_{\mu}\langle J^{\mu}_a(p) \phi (y) \rangle&=& e^{ipy}  \langle \delta_a\phi(y) \rangle \nonumber\\
p_{\mu}\langle J^{\mu}_a(p) e^{-ipy} \phi (y) \rangle&=&   \langle \delta_a\phi(y) \rangle, 
\label{1.9}
\end{eqnarray}
where we have identified the Fourier transform of the current as $J^{\mu}_a(p) = \int d^Dx e^{ipx} J^{\mu}_a(x)$.
Then, integrating both sides over $y$, 		
\begin{equation}
p_{\mu}\langle J^{\mu}_a(p) \phi (-p) \rangle=  \int d^Dy \langle \delta_a\phi(y) \rangle= \langle \delta_a\phi(p=0) \rangle.
\label{1.10}
\end{equation}		
The object $\langle \delta_a\phi(p=0) \rangle$ in the right hand side is the order parameter characterizing the possible phases of the theory \cite{Beekman:2019pmi}. The symmetric phase corresponds to $\langle \delta_a\phi(p=0) \rangle=0$, whereas $ \langle \delta_a\phi(p=0) \rangle \neq 0$ implies spontaneous symmetry breaking. In the broken phase, therefore, the correlation function $\langle J^{\mu}_a(p) \phi (-p) \rangle$ must have a pole at zero momentum, 
\begin{equation}
\langle J^{\mu}_a(p) \phi (-p) \rangle \sim \frac{p^{\mu}}{p^2}.
\label{1.11}
\end{equation}
This, in turn, signals the presence of massless physical excitations in the spectrum. These excitations are the Goldstone bosons. 
		


\section{Rephrasing Ordinary Symmetries in Terms of Topology}\label{Sec3}

Now let us discuss the ordinary symmetries in a language that is useful in generalizing to the case of higher-form symmetries. Differential forms are helpful here and we have summarized some of their properties in the Appendix \ref{diff}.

In terms of differential forms, the  Noether current can be thought as the components of a 1-form,
\begin{equation}
J=J_{\mu}dx^{\mu},
\label{2.6}
\end{equation}
whereas its Hodge dual, 
\begin{equation}
*J\equiv \frac{1}{(D-1)!} J_{\mu}\, \epsilon^{\mu}{}_{\mu_1\ldots\mu_{D-1}} dx^{\mu_1}\wedge\cdots \wedge dx^{\mu_{D-1}},
\label{2.7}
\end{equation}
is a $(D-1)$-form. For simplicity, from now on we omit the index $a$ of the current. The conservation law in \eqref{0.4} is written as
\begin{equation}
d*J=0.
\label{2.8}
\end{equation}
This means that $*J$ is a closed form. In components the left hand side reads
\begin{eqnarray}
d*J&=&\frac{1}{(D-1)!} (\partial_{\alpha}J_{\mu})\epsilon^{\mu}{}_{\mu_1\ldots\mu_{D-1}} dx^{\alpha}\wedge dx^{\mu_1}\wedge\cdots \wedge dx^{\mu_{D-1}}\nonumber\\
&=&\frac{1}{(D-1)!} (\partial_{\alpha}J_{\mu})\epsilon^{\mu}{}_{\mu_1\ldots\mu_{D-1}}  \epsilon^{\alpha\mu_1\ldots \mu_{D-1}} dx^0\wedge \cdots \wedge dx^{D-1}\nonumber\\
&=&(-1)^{D-1}(\partial_{\mu}J^{\mu})dx^0\wedge \cdots \wedge dx^{D-1},
\label{2.9}
\end{eqnarray}
where we have used the convention $\epsilon^{01\ldots D-1}\equiv +1$, and
\begin{equation}
\epsilon^{\mu}{}_{\mu_1\ldots\mu_{D-1}}  \epsilon^{\alpha\mu_1\ldots \mu_{D-1}}=(-1)^{D-1} (D-1)! \,\eta^{\mu\alpha},
\label{2.10}
\end{equation} 
with the metric $\eta^{\mu\nu}=(+,-,-,\ldots,-)$. In the Euclidean, we simply drop out the factor $(-1)^{D-1}$ on the right hand side and replace the metric $\eta^{\mu\nu}$ with $\delta^{\mu\nu}$. In this case, the conservation law is simply
\begin{equation}
d*J=\partial_{\mu}J^{\mu}dx^0\wedge \cdots \wedge dx^{D-1}.
\label{2.10a}
\end{equation}

\subsection{Assigning Topological Meaning to Charges}

\begin{figure}
	\centering
	\includegraphics[scale=.7,angle=90]{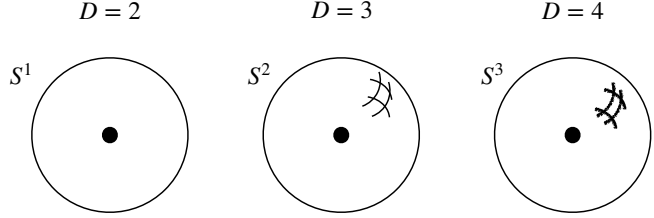}
	\caption{Link of spheres $S^{D-1}$ and a point.}
	\label{fig0}
\end{figure}

The Noether charges \eqref{0.5} can be written as an integral over a closed $(D-1)$-dimensional submanifold $\Sigma$:
\begin{eqnarray}
Q(\Sigma)&=&\int_{\Sigma} *J=\int_{\Sigma} \frac{1}{(D-1)!} J_{\mu} \,\epsilon^{\mu}{}_{\mu_1\ldots\mu_{D-1}}dx^{\mu_1}\wedge\cdots \wedge dx^{\mu_{D-1}}.
\label{2.11}
\end{eqnarray}
To see this, we consider the Euclidean spacetime, where time and space coordinates are treated on an equal foot. Then, more generally, we can  integrate the Ward identity in \eqref{1.6}  over a $D$-dimensional region $\Omega_\Sigma$ so that its boundary is the $(D-1)$-dimensional submanifold  $\Sigma$, i.e., $\partial\Omega_{\Sigma}=\Sigma$. With this, the left hand side of \eqref{1.6} becomes
\begin{eqnarray}
\int_{\Omega_{\Sigma}}\langle  d*J \phi (y)\rangle &=&  \int_{\Sigma} \langle *J \phi(y) \rangle\nonumber\\
&=& \langle Q(\Sigma) \phi(y)\rangle, 
\end{eqnarray}
where in the first line we have used the Stokes theorem \eqref{2.5a}. According to \eqref{1.6}, we obtain
\begin{equation}
\langle Q(\Sigma) \phi(y)\rangle = -i \int_{\Omega_{\Sigma}} d^{D}x \delta^{(D)}(x-y) \langle \delta_a\phi(y) \rangle.
\label{2.12}
\end{equation}	
In the right hand side we identify $ \int_{\Omega_{\Sigma}} d^{D}x \delta(x-y)$ as  the intersection number of $\Omega_{\Sigma}$ and $y$. This, in turn, is equal the link number of $\Sigma$ and $y$,
\begin{equation}
\text{Link}(\Sigma,y)= \int_{\Omega_{\Sigma}} d^{D}x \delta^{(D)}(x-y),
\label{2.13}
\end{equation}	
which is 0 or 1 depending whether $y$ is inside the region $\Omega_{\Sigma}$ or not. With this we can write \eqref{2.12} as
\begin{equation}
\langle Q(\Sigma) \phi(y)\rangle =-i \, \text{Link}(\Sigma,y)  \langle \delta\phi(y) \rangle.
\label{2.14}
\end{equation}
The link number defined in \eqref{2.13} is clearly topological since it is unaffected by deformations of the surface $\Sigma$ as long as the deformations do not cross the point $y$. We can understand that the charge $Q(\Sigma)$ is also a topological invariant by considering a deformation of the original region $\Omega_{\Sigma}$ to $\Omega_{\Sigma'}'=\Omega_{\Sigma}\cup \Omega_0$, such that $y$ does not belong to $\Omega_0$. This implies
\begin{eqnarray}
Q(\Sigma+\partial\Omega_0)=\int_{\Omega_{\Sigma}\cup \Omega_0}\langle  d*J \phi (y)\rangle &=&  \int_{\Omega_{\Sigma}}\langle  d*J \phi (y)\rangle+ \int_{ \Omega_0}\langle  d*J \phi (y)\rangle\nonumber\\
&=&  \int_{\Omega_{\Sigma}}\langle  d*J \phi (y)\rangle = Q(\Sigma),
\label{2.15}
\end{eqnarray}
where, as $y \not\in \Omega_0$, we have set $d*J=0$ inside the correlator in the last term of the first line. Therefore, the conservation law is translated into the fact that the operator $Q(\Sigma)$ is topological. We notice that for any spacetime dimensionality $D$, a point can always link with a closed $D-1$ manifold like $S^{D-1}$ that surrounds it, as shown in Fig. \ref{fig0}.

We can also write the finite form of the relation \eqref{2.14}, namely,
\begin{equation}
\langle U(g,\Sigma) \phi(y)\rangle = R(g) \langle \phi (y)\rangle, 
\label{2.16}
\end{equation}
if $y$ and $\Sigma$ are linked. $U(g,\Sigma)$ is the unitary topological operator associated with the symmetry group $g$ and $R$ stands for the representation in which the fields transform,
\begin{equation}
U(g,\Sigma)=e^{i \alpha_a Q_a}~~~\text{and}~~~ R(g)= e^{\alpha_a t_a},
\label{2.17}
\end{equation}  
with $t_a$ corresponding to the generators in the representation that $\phi$ belongs. For infinitesimal parameters $\alpha_a$ this recovers immediately the relation in \eqref{2.14}, with the identification $\delta_a\phi\equiv t_a\phi$. We refer to this as a $0$-form symmetry, in the sense that the charged objects under the symmetry are local operators $\phi(y)$ supported in a point, i.e., in a $0$-dimensional region. Equivalently, the parameter of transformation is a closed 0-form, which is simply a constant.

\begin{figure}
	\centering
	\includegraphics[scale=.77,angle=90]{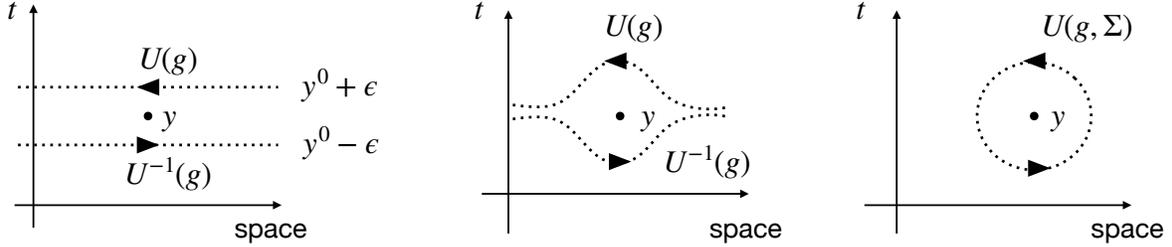}
	\caption{Sequence of deformations to associate a closed surface $\Sigma$ to $U(g,\Sigma)$.}
	\label{def}
\end{figure}

For discrete symmetries, even though a conserved charge does not exist, we can nevertheless define a topological unitary operator precisely as in \eqref{2.16}. In fact, consider the unitary operator for a discrete symmetry, $U(g)$ (with no parameter involved), and the corresponding  action over a local operator,
\begin{equation}
\langle U(g) \phi(y) U^{-1}(g)\rangle = R(g)\langle\phi(y)\rangle.
\label{2.18}
\end{equation}
In this relation, the operator $U(g)$ is interpreted as defined at a time $y^0+\epsilon$ and the operator $U^{-1}(g)$ at the time $y^0-\epsilon$. The equal-time is understood as the limit $\epsilon\rightarrow 0$. Furthermore, we can associate a spatial slice with the operator $U(g)$. Next we assume that $[U(g),P_{\mu}]$, where $P_{\mu}$ is the generator of spacetime translations. This implies that the spacetime region associated with $U(g)$ can be continuously deformed into a closed one, through the sequence of steps shown in Fig. \ref{def}. In this case, the left hand side of \eqref{2.18} can be written as
\begin{equation}
\langle U(g) \phi(y) U^{-1}(g)\rangle = \langle U(g,\Sigma)\phi(y)\rangle,
\end{equation}
when $y$ and $\Sigma$ are linked. This leads to the conclusion that the relation \eqref{2.16} is also valid for discrete symmetries. 

The above construction provides an interesting perspective on symmetries in relativistic theories, namely,
\begin{equation}
\text{Symmetry Generator}~~~ \Leftrightarrow ~~~\text{Topological Operator},
\end{equation}
with the charged operators corresponding to the objects with nontrivial link with the topological operator. This perspective is illuminating and provides a natural way to generalize the notion of symmetry to the case of higher-forms.


\section{Higher-Form Symmetries}\label{Sec4}		
		
Now we generalize the previous discussion to the case of a $1$-form symmetry\footnote{We will discuss the general case of $q$-form symmetries a little later.}. To this, it is useful to reconsider the case of an ordinary symmetry. It is referred to as a 0-form symmetry, in the sense that the involved parameter $\xi$ is a closed 0-form, i.e., $d \xi=0$ (since it corresponds to a global symmetry). In terms of differential forms, the expression \eqref{0.2} becomes
\begin{equation}
\delta S = \int_{\mathcal{M}^{(D)}} *J \wedge d\xi, 
\label{3.1}
\end{equation}
where we have promoted $\xi$ to a local parameter, i.e., $d\xi$  is no longer closed in this expression. Upon an integration by parts, and using the equation of motion, the above expression leads to 
\begin{equation}
d*J=0.
\label{3.2}
\end{equation}

Now let us study a generalization of the above setting by considering a global symmetry whose parameter is a closed 1-form $\xi_1=\xi_{\mu}dx^{\mu}$, namely, $d\xi_1=0$. We emphasize that the global nature of the transformation is translated into the {\it flatness} condition on the parameter, $\partial_{\mu}\xi_{\nu}-\partial_{\nu}\xi_{\mu}=0$. The analogue of expression \eqref{3.1}, obtained by removing the closeness condition on the parameter $\xi_1$, is
\begin{equation}
	\delta S = \int_{\mathcal{M}^{(D)}} *J \wedge d\xi_1, 
	\label{3.3}
\end{equation}
where now $*J$ is a $(D-2)$-form or, equivalently, $J$ is a 2-form, 
\begin{equation}
J=\frac{1}{2!}J_{\mu\nu}dx^{\mu}\wedge dx^{\nu}.
\label{3.4}
\end{equation}
In components, the conservation law $d*J=0$ reads
\begin{equation}
\partial_{\mu}J^{\mu\nu}=0,~~~\text{with}~~J^{\mu\nu}=-J^{\nu\mu}.
\label{3.5}
\end{equation}
As $*J$ is a $(D-2)$-form, in analogy to what we did in the ordinary case (eq. \eqref{2.11}), we can define the charge on a closed $\Sigma_{D-2}$ submanifold, 
\begin{equation}
Q(\Sigma_{D-2})\equiv \int_{\Sigma_{D-2}} *J.	
\label{3.6}
\end{equation}
Next we look for the charged objects under this symmetry operator or, in other words, the objects that possess nontrivial link with $\Sigma_{D-2}$.

\subsection{Charged Operators}

We wish to understand now the nature of the charged objects under the 1-form symmetry. To address this, it is convenient to reconsider the ordinary case but under a perspective that is helpful in generalizing. We remember the case of a 0-form symmetry, where under an infinitesimal global transformation characterized by the constant parameter $\xi$, a local operator transforms as
\begin{equation}
\phi(x) \rightarrow \phi'(x)=\phi(x) +\xi \delta\phi(x).
\label{3.7}
\end{equation} 
In other words, we have assumed from the beginning  that charged operators under the symmetry are the local operators, i.e., operators with support in a 0-dimensional region of spacetime. 

So the question is whether there is a unambiguous way to see that the charged objects are in fact local operators. From the perspective of the Hilbert space, the transformation in \eqref{3.7} can be interpreted as coming from the actuation of the charge operator defined over a spatial $D-1= d $ dimensional slice (without boundaries) at a fixed time. Then we can use the notion of Poincaré duality to associate a form with a manifold. More precisely, the Poincaré duality provides a way to associate a $(d-p)$-form with a $p$-dimensional manifold. The components of the Poincaré dual $(d-p)$-form are defined as
\begin{equation}
	\xi_{i_{p+1}\ldots i_d}(x)\equiv \frac{1}{p!}\int_{\Sigma_{p}} \epsilon_{i_1\ldots i_p i_{p+1}\ldots i_{d}}\delta^{(d)}(\vec{x}-\vec{y}) dy^{i_{1}}\wedge \cdots \wedge dy^{i_{p}}. 
	\label{3.8}
\end{equation}
The paramater of the global symmetry is then identified as the $(d-p)$-form $\xi_{d-p}(\Sigma_{p})$ constructed from the submanifold of dimension $p$. As we shall see, this automatically ensures that $\xi_{d-p}$ is closed, which is the condition for the symmetry to be global.  

In this context, the parameter of the transformation of an ordinary symmetry is identified, up to a constant factor, as the Poincaré dual of the $\Sigma_{d}$ spatial manifold, which is just a 0-form constant. This follows
simply by setting $p=d$ in \eqref{3.8},
\begin{equation}
\xi(x)=\frac{1}{d!}\int_{\Sigma_{d}} \epsilon_{i_1\ldots i_{d}}\delta^{(d)}(\vec{x}-\vec{y}) dy^{i_{1}}\wedge \cdots \wedge dy^{i_{d}}=1.
\label{3.9}
\end{equation}
This means that the parameters of ordinary symmetries are closed 0-forms, which are supported on 0-dimensional regions (points) of the manifold. Accordingly, they can be associated with the transformation of similar objects, namely, objects also supported on 0-dimensional regions of the manifold - the local operators.

Now it is easy to see how this picture generalizes for higher-form symmetries. If we have a submanifold of dimension $p=d-1$, then the Poincaré dual is a 1-form $\xi_1(\Sigma_{d-1})$ with components 
\begin{equation}
\xi_{i_d}(x)= \frac{1}{(d-1)!}\int_{\Sigma_{d-1}} \epsilon_{i_1\ldots i_{d-1} i_{d}}\delta^{(d)}(\vec{x}-\vec{y}) dy^{i_{1}}\wedge \cdots \wedge dy^{i_{d-1}}.
\label{3.10}
\end{equation}  
Therefore, the objects that are charged under the 1-form symmetry are operators with support along a line - the line operators.  So given an operator supported along a line $\mathcal{C}$, the parameter of the transformation (up to a constant factor\footnote{We can think that the parameter of the transformation is absorbed into the Poincaré dual $\xi_1$.}) is 
\begin{equation}
\int_{\mathcal{C}} \xi_1(\Sigma_{d-1})= \int_{\mathcal{C}} \xi_i dx^i.
\label{3.11}
\end{equation}
More explicitly, the infinitesimal transformation of a line operator reads
\begin{equation}
W[\mathcal{C}]\rightarrow W[\mathcal{C}]'=W[\mathcal{C}]+ \int_{\mathcal{C}} \xi_1(\Sigma_{d-1}) \delta W[\mathcal{C}],
\label{3.12}
\end{equation}
It is worth to emphasize that, even when the line $\mathcal{C}$ is closed, the integral $ \int_{\mathcal{C}} \xi_1(\Sigma_{d-1})$ may not vanish in spite of the fact that $d\xi_1=0$. Naively, if we use the Stokes theorem, we could convert the line integral to a surface integral that has $\mathcal{C}$ as the boundary, $\int_{\mathcal{C}} \xi_1(\Sigma_{d-1})=\int_{S} d\xi_1$. However, this may not be true because of eventual topological obstructions to employing the Stokes theory caused by $\Sigma_{d-1}$, when $\mathcal{C}$ and $\Sigma_{d-1}$ intersect. We shall discuss this in a moment.

Now we can check that $\xi_1=\xi_i dx^i$ is closed because $\Sigma_{d-1}$ has no boundary. In fact, we first write it as
\begin{equation}
\xi_{i}(x)= \int_{\Sigma_{d-1}}  (d\Sigma^{d-1})_{i}(y)\delta^{(d)}(\vec{x}-\vec{y}),
\label{3.13}
\end{equation}
where $(d\Sigma^{d-1})_{i_d}(y)$ is the oriented integration element over the subspace $\Sigma_{d-1}$ with coordinates $y$. Then we take the exterior derivative
\begin{eqnarray}
d\xi_1&=& \partial^x_k \xi_i(x) dx^k \wedge dx^i\nonumber\\
&=&  \int_{\Sigma_{d-1}}  (d\Sigma^{d-1})_{i}(y) \partial^x_k \delta^{(d)}(\vec{x}-\vec{y}) dx^k \wedge dx^i\nonumber\\
&=&  -\int_{\Sigma_{d-1}}  (d\Sigma^{d-1})_{i}(y) \partial^y_k \delta^{(d)}(\vec{x}-\vec{y}) dx^k \wedge dx^i.
\label{3.14}
\end{eqnarray}
Suppose that $\Sigma_{d-1}$ dimensional manifold is the infinite space (without boundary) associated with the directions $x^{1},\ldots, x^{d-1}$, so that the element $(d\Sigma^{d-1})_{i_d}(y)$ is oriented along direction $x^{d}$. In this case, the above expression becomes
\begin{eqnarray}
d\xi_1&=&-\int_{\Sigma_{d-1}}  (d\Sigma^{d-1})_{d}(y) \partial^y_k \delta^{(d)}(\vec{x}-\vec{y}) dx^k \wedge dx^d\nonumber\\
&=&-\int_{\Sigma_{d-1}}  (d\Sigma^{d-1})_{d}(y) \vec{\nabla}_y \cdot \left( \delta^{(d)}(\vec{x}-\vec{y}) d\vec{x} \wedge dx^d\right)=0,
\label{3.15}
\end{eqnarray}
i.e., it vanishes upon using the divergence theorem and taking into account that $\Sigma_{d-1}$ has no boundaries.

We can generalize the manifold in which the charge is defined $\Sigma_{d-1}$ and consider instead $\Sigma_{D-2}$ as an arbitrary closed manifold in spacetime. We can also generalize the notion of a line operator (an operator that acts on the Hilbert space at fixed time) to a {\it defect} line, which is an operator also extended along the time direction\footnote{In relativistic theories this distinction is tenuous.}. In this case, we have a more general symmetry transformation acting on a defect line,
 \begin{equation}
 W[\mathcal{C}]\rightarrow W[\mathcal{C}]'=W[\mathcal{C}]+ \int_{\mathcal{C}} \xi_1(\Sigma_{D-2}) \delta W[\mathcal{C}],
 \label{3.16}
 \end{equation} 
where now the line $\mathcal{C}$ is extended along the time direction. 

From this expression we can derive the corresponding Ward identities. Let us consider the correlation function involving a single defect,
\begin{eqnarray}
\langle W[\mathcal{C}]\rangle &=& \int \mathcal{D}\phi W[\mathcal{C}] e^{iS[\phi]}\nonumber\\
&=&  \int \mathcal{D}\phi' W'[\mathcal{C}] e^{iS[\phi']}\nonumber\\
&=& \int \mathcal{D}\phi \left(W[\mathcal{C}]+ \int_{\mathcal{C}} \xi_1(\Sigma_{D-2}) \delta W[\mathcal{C}]\right) \left(1+i \delta S \right) e^{iS[\phi']},
\label{3.17}
\end{eqnarray}
with $\delta S$ given in \eqref{3.3}, which in components reads  
\begin{equation}
\delta S = \int d^Dx J^{\mu\nu} \partial_{\mu}\xi_{\nu} = - \int d^Dx \,\xi_{\nu} \,\partial_{\mu}J^{\mu\nu}.
\label{3.18}
\end{equation}
Relation \eqref{3.17} implies
\begin{eqnarray}
i\int d^Dx \xi_{\nu}(x) \langle \partial_{\mu}J^{\mu\nu} W[\mathcal{C}]  \rangle &=& \int_{\mathcal{C}} dy^{\nu} \xi_{\nu}(y) \langle \delta W[\mathcal{C}]\rangle\nonumber\\
&=& \int d^Dx \,\xi_{\nu}(x)  \int_{\mathcal{C}} \delta^{(D)}(x-y) dy^{\nu}  \langle \delta W[\mathcal{C}]\rangle.
\label{3.19}
\end{eqnarray}
By factorizing $\xi_{\nu}(x)$, it follows that 
\begin{equation}
	\langle \partial_{\mu}J^{\mu\nu}(x) \,W[\mathcal{C}] \rangle = -i \int_{\mathcal{C}} dy^{\nu} \delta^{(D)}(x-y) \langle \delta W[\mathcal{C}]\rangle,
	\label{3.20}
\end{equation}
which is the Ward identity for a single line defect. 

Now we are ready to explore further consequences of the 1-form symmetry.


\subsection{Case Study: 1-Form Symmetry in $D=2$}

\begin{figure}
	\centering
	\includegraphics[scale=1,angle=90]{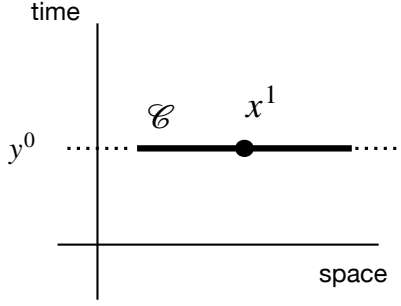}
	\caption{Intersection of the line $\mathcal{C}$ with the spatial point $x^1$.}
	\label{fig1}
\end{figure}
		
Let us start with the simplest situation of a 1-form symmetry in $D=2$ spacetime dimensions. In this case, the conservation law is quite simple,		
\begin{equation}
\partial_{0} J^{01}=0~~~\text{and}~~~\partial_{1} J^{10}=0,
\end{equation}		
implying that $J^{01}$ is conserved in a zero dimensional subspace, since it does not need to be integrated along the $x^1$-direction. The presence of a line operator may lead to the violation of the conservation law according to the Ward identity in \eqref{3.20}, which can be written more explicitly as
\begin{equation}
\langle \partial_{0}J^{01}(x) \,W[\mathcal{C}] \rangle = -i \int_{\mathcal{C}} dy^{1} \delta^{(2)}(x-y) \langle \delta W[\mathcal{C}]\rangle.
\label{3.21}
\end{equation}
Assuming that $W[\mathcal{C}]$ is a true line operator, i.e., that the curve $\mathcal{C}$ is extended along the spatial direction at fixed time $y^0$, and integrating \eqref{3.21} along $x^0$ from $y^0-\epsilon$ to $y^0+\epsilon$, we get
\begin{equation}
\langle [J^{01},W[\mathcal{C}]]\rangle = -i \int_{\mathcal{C}}dy^1 \delta(x^1-y^1) \langle \delta W[\mathcal{C}] \rangle. 
\label{3.22}
\end{equation}
The integral $\int_{\mathcal{C}}dy^1 \delta(x^1-y^1) $ is the intersection of the line $\mathcal{C}$ with the 0-dimensional submanifold in which the charge is defined, namely, the spatial point $x^1$. Whenever they intersect, as shown in Fig. \ref{fig1}, this integral is equal to one, and thus the Ward identity shows how the line operator transforms under the symmetry.

\begin{figure}
	\centering
	\includegraphics[scale=0.75,angle=90]{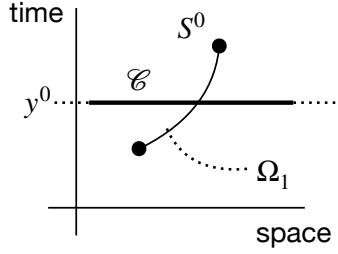}
	\caption{Link of the curve $\mathcal{C}$ with the $S^0$ sphere.}
	\label{fig2}
\end{figure}

\begin{figure}
	\centering
	\includegraphics[scale=0.9,angle=90]{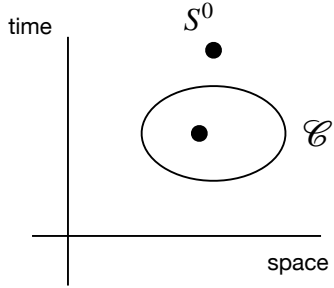}
	\caption{Topological meaning of $\text{Link}(S^0,\mathcal{C})$.}
	\label{fig3}
\end{figure}

Now we can take a different perspective and consider the 0-dimensional region where the charge is defined as a sphere $S^0$, which consists of two points, as illustrated in Fig. \ref{fig2}.
We then integrate the Ward identity on a 1-dimensional space $\Omega_1$, whose boundary is $\partial\Omega_1=S^0$,
\begin{equation}
	\int_{\Omega_1}(d\Omega_1)_{\nu} \langle \partial_{\mu}J^{\mu\nu}(x) \,W[\mathcal{C}] \rangle = -i \int_{\Omega_1}(d\Omega_1)_{\nu} \int_{\mathcal{C}} dy^{\nu} \delta^{(D)}(x-y) \langle \delta W[\mathcal{C}]\rangle,
	\label{3.23}
\end{equation}
where $(d\Omega_1)_{\nu} $ is the oriented element of integration on $\Omega_1$. With this, \eqref{3.23} can be written as
\begin{equation}
\langle J^{01}(S^0) W[\mathcal{C}]\rangle = -i \,\text{Link}(S^0,\mathcal{C})  \langle \delta W[\mathcal{C}]\rangle.
\label{3.24}
\end{equation}
On the left hand side, we have the charge defined on $S^0$, whereas on the right hand side we have the intersection number of $\Omega_1$ and $\mathcal{C}$, which is equal to the link number of $\mathcal{C}$ and $S^{0}$,
\begin{equation}
\int_{\Omega_1}(d\Omega_1)_{\nu} \int_{\mathcal{C}} dy^{\nu} \delta^{(D)}(x-y)= \text{Link}(S^0,\mathcal{C}).
\label{3.25}
\end{equation}
As long as the curve $\mathcal{C}$ is infinitely extended or closed, the $\text{Link}(S^0,\mathcal{C})$ has a topological meaning, as shown in Fig. \ref{fig3}. Relation \eqref{3.24} is the analog of relation \eqref{2.14}.

\subsection{Case Study: 1-Form Symmetry in $D=3$}		

Now we consider a 1-form symmetry in $D=3$ spacetime dimensions.  The conservation laws are
\begin{equation}
\partial_{\mu}J^{\mu\nu}=\partial_{0} J^{0\nu}+\partial_{1} J^{1\nu}+\partial_{2} J^{2\nu}=0,
\label{3.26}
\end{equation}
which give rise to 
\begin{equation}
\partial_{0}J^{01}+\partial_{2} J^{21}=0~~~\text{and}~~~\partial_{0}J^{02}+\partial_{1} J^{12}=0.
\label{3.27}
\end{equation}
It follows that the charges,
\begin{equation}
Q^1=\int dx^2 J^{01}~~~\text{and}~~~Q^2=\int dx^1 J^{02},
\label{3.28}
\end{equation}
defined in spatial 1-dimensional submanifolds, are conserved. It is also interesting to notice that the charge $Q^1$ does not depend on the coordinate $x^1$, even though it is integrated only along direction $x^2$. This follows from the conservation law \eqref{3.26}, by setting $\nu=0$, 
\begin{equation}
\partial_{1} J^{10}+\partial_{2} J^{20}=0, 
\end{equation}
which implies that 
\begin{equation}
\frac{\partial Q^1 }{\partial x^1} =0.
\end{equation}
The same reasoning applies to the charge $Q^2$, implying that it is independent on $x^2$. This is in contrast with the subsystem symmetries mentioned in the Introduction, where  the conserved charges carry dependence on the coordinates of certain submanifolds, so that there are actually an infinite number of charges (in the continuum limit). In general, this implies that the states of the spectrum are infinitely degenerated.

\begin{figure}
	\centering
	\includegraphics[scale=.75,angle=90]{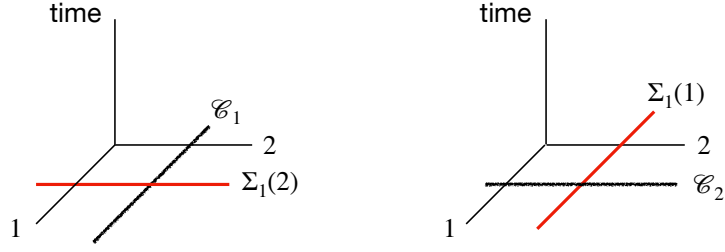}
	\caption{Intersections of lines at fixed time.}
	\label{fig5}
\end{figure}

The violation of the conservation law due to the presence of a line operator is dictated by the Ward identity
\begin{equation}
\langle \partial_{\mu}J^{\mu\nu}(x) \,W[\mathcal{C}] \rangle = -i \int_{\mathcal{C}} dy^{\nu} \delta^{(3)}(x-y) \langle \delta W[\mathcal{C}]\rangle.
\label{3.29}
\end{equation}
The quantum counterpart of equations in \eqref{3.27} are
\begin{equation}
\langle \partial_{0}J^{01}(x) \,W[\mathcal{C}] \rangle +\langle \partial_{2}J^{21}(x) \,W[\mathcal{C}] \rangle  = -i \int_{\mathcal{C}} dy^{1} \delta^{(3)}(x-y) \langle \delta W[\mathcal{C}]\rangle
\label{3.30}
\end{equation}
and 
\begin{equation}
\langle \partial_{0}J^{02}(x) \,W[\mathcal{C}] \rangle +\langle \partial_{1}J^{12}(x) \,W[\mathcal{C}] \rangle  = -i \int_{\mathcal{C}} dy^{2} \delta^{(3)}(x-y) \langle \delta W[\mathcal{C}]\rangle.
\label{3.31}
\end{equation}
Integrating these expressions on 1-dimensional spatial submanifolds $\Sigma_1(1)$ and $\Sigma_1(2)$ and over the time coordinate $x^0$ from $y^0-\epsilon$ to $y^0+\epsilon$, leads to 
\begin{equation}
\langle [Q^1, W[\mathcal{C}_1]] \rangle = -i \int_{\Sigma_1(2)} dx^2  \int_{\mathcal{C}_1} dy^{1} \delta^{(2)}(\vec{x}-\vec{y}) \langle \delta W[\mathcal{C}_1]\rangle
\label{3.32}
\end{equation}
and
\begin{equation}
	\langle [Q^2, W[\mathcal{C}_2]] \rangle = -i \int_{\Sigma_1(1)} dx^1  \int_{\mathcal{C}_2} dy^{2} \delta^{(2)}(\vec{x}-\vec{y}) \langle \delta W[\mathcal{C}_2]\rangle.
	\label{3.33}
\end{equation}
The integrals on the right hand side of these expressions correspond to the intersection of the lines, which are shown in Fig. \ref{fig5}.

\begin{figure}
	\centering
	\includegraphics[scale=.7,angle=90]{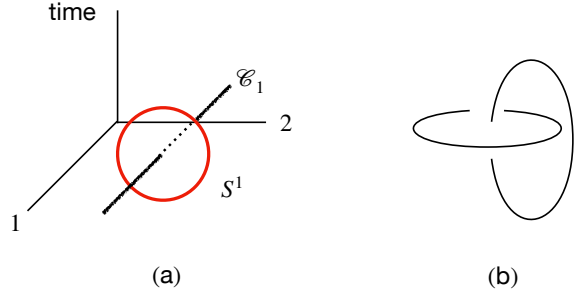}
	\caption{(a) Link of curves. (b) Link of two generic closed curves in $D=3$. }
	\label{fig11}
\end{figure}

By following the previous strategy in assigning a topological meaning to the conservation laws, instead of integrating on 1-dimensional submanifolds $\Sigma_1(1)$ and $\Sigma_1(2)$ at fixed time, we can consider a closed general curve in spacetime like a $S^1$. In this case, we integrate both sides of the Ward identity \eqref{3.29} over a 2-dimensional manifold $\Omega_2$ whose boundary is $\partial\Omega_2=S^1$. Denoting the oriented integration element on $\Omega_2$ by $(d\Omega_2)_{\nu}$, it follows that
\begin{equation}
	\int_{\Omega_2} (d\Omega_2)_{\nu}  \langle \partial_{\mu}J^{\mu\nu}(x) \,W[\mathcal{C}] \rangle = -i \int_{\Omega_2} (d\Omega_2)_{\nu}  \int_{\mathcal{C}} dy^{\nu} \delta^{(3)}(x-y) \langle \delta W[\mathcal{C}]\rangle,
	\label{3.34}
\end{equation}
which can be expressed as
\begin{equation}
\langle Q(S^1) W[\mathcal{C}] \rangle = -i \text{Link}(S^1,\mathcal{C})  \langle \delta W[\mathcal{C}]\rangle.
\label{3.35}
\end{equation}
The link is shown in Fig. \ref{fig11}(a).	Fig. \ref{fig11}(b) illustrates the link of two generic closed curves. 	

%



		
\subsection{Case Study: 1-Form Symmetry in $D=4$}				
		
Next we consider a 1-form symmetry in $D=4$ spacetime dimensions. The conservation laws are
\begin{equation}
	\partial_{\mu}J^{\mu\nu}=\partial_{0} J^{0\nu}+\partial_{1} J^{1\nu}+\partial_{2} J^{2\nu}+\partial_{3}J^{3\nu}=0.
	\label{3.36}
\end{equation}
From these conservation laws it follows that the charges,
\begin{equation}
	Q^1=\int dx^2 dx^3 J^{01}, ~~~Q^2=\int dx^1 dx^3 J^{02},~~~\text{and}~~~Q^3=\int dx^1 dx^2 J^{03},
	\label{3.38}
\end{equation}
defined in spatial 2-dimensional submanifolds, are conserved. As discussed in the case $D=3$, we emphasize again that $Q^1$ is independent on $x^1$, $Q^2$ is independent of $x^2$, and $Q^3$ is independent on $x^3$, which follows from \eqref{3.36} with $\nu=0$,
\begin{equation}
\partial_{1} J^{10}+\partial_{2} J^{20}+\partial_{3}J^{30}=0.
\label{3.39}
\end{equation}		
By integrating on the respective 2-dimensional subspaces we get the above conclusion. 

\begin{figure}
	\centering
	\includegraphics[scale=.7,angle=90]{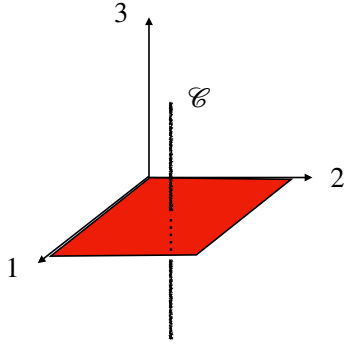}
	\caption{Intersection between a two-dimensional spatial slice (in red) and a curve $\mathcal{C}$ at a fixed time.}
	\label{fig7}
\end{figure}

Now we consider the Ward identity \eqref{3.20} in $D=4$,
\begin{equation}
	\langle \partial_{\mu}J^{\mu\nu}(x) \,W[\mathcal{C}] \rangle = -i \int_{\mathcal{C}} dy^{\nu} \delta^{(4)}(x-y) \langle \delta W[\mathcal{C}]\rangle.
	\label{3.40}
\end{equation}
Choosing, say, $\nu=3$, we have
\begin{equation}
	\langle \partial_{0}J^{03}(x) \,W[\mathcal{C}] \rangle+	\langle \partial_{1}J^{13}(x) \,W[\mathcal{C}] \rangle +	\langle \partial_{2}J^{23}(x) \,W[\mathcal{C}] \rangle  = -i \int_{\mathcal{C}} dy^{3} \delta^{(4)}(x-y) \langle \delta W[\mathcal{C}]\rangle.
	\label{3.41}
\end{equation}
Integrating both sides over $\int_{y^0-\epsilon}^{{y^0+\epsilon}} dx^0 dx^1 dx^2$, it follows that
\begin{equation}
\langle [Q^3, W[\mathcal{C}_1]] \rangle = -i \int dx^1 dx^2 \int_{\mathcal{C}} dy^3 \delta^{(3)}(\vec{x}-\vec{y})\langle \delta W[\mathcal{C}]\rangle,
\label{3.42}
\end{equation}
where the integrals in the right hand side correspond to the intersection between the two-dimensional surface in the plane $x^1$-$x^2$ and the line $\mathcal{C}$ in the direction $3$, as shown in Fig. \ref{fig7}. The same reasoning follows for curves along the remaining directions. 

\begin{figure}
	\centering
	\includegraphics[scale=.6,angle=90]{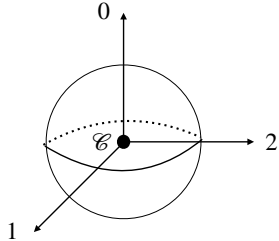}
	\caption{The line $\mathcal{C}$ appears as a single point at the origin, which is linked with $S^2$.}
	\label{fig8}
\end{figure}

To assign a topological meaning to the charge we go back to the Ward identity \eqref{3.40} and integrate both sides over a region $\Omega_3$, with $\partial \Omega_3=S^2$,
\begin{equation}
\int_{\Omega_3} (d\Omega_3)_{\nu}	\langle \partial_{\mu}J^{\mu\nu}(x) \,W[\mathcal{C}] \rangle = -i \int_{\Omega_3} (d\Omega_3)_{\nu} \int_{\mathcal{C}} dy^{\nu} \delta^{(4)}(x-y) \langle \delta W[\mathcal{C}]\rangle.
	\label{3.43}
\end{equation}
As in the previous cases, we identify the intersection number as the link between the curve $\mathcal{C}$ and $S^2$,
\begin{equation}
\int_{\Omega_3} (d\Omega_3)_{\nu} \int_{\mathcal{C}} dy^{\nu} \delta^{(4)}(x-y) = \text{Link}(S^2,\mathcal{C}).
\label{3.44}
\end{equation}
Therefore, we can write \eqref{3.43} as
\begin{equation}
	\langle Q(S^2) W[\mathcal{C}] \rangle = -i \text{Link}(S^2,\mathcal{C})  \langle \delta W[\mathcal{C}]\rangle.
	\label{3.45}
\end{equation}
For a line oriented along direction 3, for example, the $S^2$ surface is immersed in the three-dimensional region $x^0$-$x^1$-$x^2$, as depicted in Fig. \ref{fig8}.


\subsection{Generalization: $q$-form Symmetries}

The generalization for the case of a $q$-form symmetry is straightforward. Given a $(q+1)$-form conserved current $J$, we can construct the charge by integrating the conservation law 
$d \ast J$ over a $(D-q)$-dimensional region $\Omega_{D-q}$, whose boundary is $\partial\Omega_{D-q}=\Sigma_{D-q-1}$. The Stokes theorem leads to 
\begin{equation}
\int_{\Omega_{D-q}} d \ast J = \int_{\Sigma_{D-q-1}}\ast J,
\label{3.46}
\end{equation}
so that we identify the charge as
\begin{equation}
Q(\Sigma_{D-q-1})= \int_{\Sigma_{D-q-1}}\ast J. 
\label{3.47}
\end{equation}
Given this $(D-q-1)$-dimensional manifold, the Poincaré dual enables us to associate a $D-1 - (D-q-1)=q$-form, which is then identified as the parameter of the transformation. Correspondingly, the charged objects are operators supported on a $q$-dimensional manifold. 

\section{Higher-Form Symmetries in Maxwell Theory}\label{Sec5}

The free Maxwell theory is the prototype of a physical theory with higher-form symmetries, and we wish to discuss it in detail. The action is defined in terms of a compact $U(1)$ gauge field $a$ with dimension $[a]=1$,
\begin{equation}
S[a]= \int -\frac{1}{2 e^2} f \wedge *f= \int d^Dx -\frac{1}{4 e^2} f_{\mu\nu}f^{\mu\nu},
\label{5.1}
\end{equation}
where $f_{\mu\nu}\equiv \partial_{\mu}a_{\nu}-\partial_{\nu}a_{\mu}$ and the coupling constant $e^2$ has dimension $[e^2]=4-D$. 

When we are working with a compact group $U(1)$ rather than the noncompact one ($\mathbf{R}$), the gauge field $a$ is an angular-type variable. An immediate consequence is that the $U(1)$ charges (electric) are quantized. A natural way to see this is to place the model in a lattice \cite{Polyakov:1987ez}. In the continuum, a simple way to unveil the angular nature of the gauge fields is to define the theory in a torus. In this case, we can construct large gauge transformations that compactify the gauge field. We discuss this below.

What are the observables of the theory \eqref{5.1}? The immediate answer is that they are related to the gauge-invariant objects, which can be constructed from the field strength $f_{\mu\nu}$. These are the local operators. Is it possible to construct extended gauge-invariant operators? The answer is yes - they are the Wilson line defect/operators
\begin{equation}
W_{q_e}[\mathcal{C}]\equiv \exp \left(i q_e \oint_{\mathcal{C}} a\right),
\label{5.1a}
\end{equation}
where the curve $\mathcal{C}$ must be infinitely long or closed in order to be gauge-invariant.  
The parameter $q_e \in \mathbb{Z}$ is the charge of the Wilson line. It is an integer because of the compactness of the gauge group. A simple way to see this is by considering the time direction as a $S^1$ of length $L_0$. Then, we can construct a gauge function that wraps around it, 
\begin{equation}
\Lambda = 2 \pi \frac{x^0}{L_0}.
\label{5.1b}
\end{equation}
This, in turn, leads to the compactness condition for the gauge field $a_0$,
\begin{equation}
a_0 \sim a_0 + \frac{2 \pi }{L_0}. 
\label{5.1c}
\end{equation}
Picking a Wilson \eqref{5.1a} along time direction, the requirement that is must be invariant under such large gauge transformation leads immediately to the quantization of $q_e$.

What is the physical interpretation of line operator \eqref{5.1a}? It represents the worldline of a probe charged particle, i.e., a particle that has no dynamics. We can see this by considering the expectation value of the Wilson loop
\begin{equation}
\langle W_{q_e}[\mathcal{C}] \rangle = \int \mathcal{D}a\,  \exp \left(i q_e \oint_{\mathcal{C}} a\right) \,e^{i S[a]}.
\end{equation} 
Next we introduce a conserved current associated with a particle moving along a curve parametrized by $\vec{y}(x^0)$
\begin{equation}
J^0(x^0,\vec{x})= q_e \delta^{(d)}(\vec{x}-\vec{y}(x^0)) ~~~\text{and}~~~\vec{J}(x^0,\vec{x})= q_e \frac{d\vec{y}(x^0)}{dx^0} \delta^{(d)}(\vec{x}-\vec{y}(x^0)), 
\end{equation}
which can be written as
\begin{equation}
J^{\mu}(x^0,\vec{x})=q_e \frac{dy^{\mu}(x^0)}{dx^0} \delta^{(d)}(\vec{x}-\vec{y}(x^0)),
\end{equation}
with $y^0=x^0$. In this way, the Wilson line can be written as
\begin{eqnarray}
\exp \left(i q_e \oint_{\mathcal{C}} dy^{\mu}a_{\mu}(y)\right)&=& \exp \left(i \int dx^0 \frac{dy^{\mu}(x^0)}{dx^0}a_{\mu}(x^0,\vec{y}) \right)\nonumber\\
&=&
\exp \left(i \int d^dx \delta^{(d)}(\vec{x}-\vec{y}(x^0)) \int dx^0 \frac{dy^{\mu}(x^0)}{dx^0}a_{\mu}(x^0,\vec{x})\right)\nonumber\\
&=& \exp \left(i \int d^Dx J^{\mu}a_{\mu}\right).
\end{eqnarray}
The upshot is that the expectation value of the Wilson loop corresponds simply to coupling the theory with non-dynamical charged matter, parametrized by the current $J^{\mu}$, 
\begin{equation}
\langle W_{q_e}[\mathcal{C}] \rangle = \int \mathcal{D}a\, e^{i S[a]+i\int d^Dx J^{\mu}a_{\mu}}.
\label{evwl}
\end{equation}

The equations of motion of the action \eqref{5.1} are
\begin{equation}
\frac{1}{e^2}d*f=0~~~\text{and}~~~ df=d*(*f)=0,
\label{5.2}
\end{equation}
which, in components, read
\begin{equation}
\frac{1}{e^2}\partial_{\mu}f^{\mu\nu}=0~~~\text{and}~~~\partial_{\mu_1}*f^{\mu_1\mu_2\ldots \mu_{D-2}}=0.
\label{5.3}
\end{equation}
In the second expression, we have defined the components of the dual field strength, which is a $(D-2)$-form and hence can be written as
\begin{equation}
*f = \frac{1}{(D-2)!} *f_{\mu_1\ldots \mu_{D-2}} dx^{\mu_1}\wedge \cdots \wedge dx^{\mu_{D-2}}.
\label{5.4}
\end{equation}
Comparing this with the definition of the dual in \eqref{2.5}, we find
\begin{equation}
*f_{\nu_1\ldots \nu_{D-2}}= \frac{1}{2!}f_{\mu_1\mu_2} \epsilon^{\mu_1\mu_2}\,_{\nu_1 \ldots \nu_{D-2}}.
\label{5.5}
\end{equation}

The equations of motion in \eqref{5.2} or \eqref{5.3} imply that the theory has two types of higher-form symmetries, namely, a $1$-form electric and a $(D-3)$-form magnetic symmetries, with the currents $J_e\equiv \frac{1}{e^2}f$ and $J_m\equiv \frac{1}{2\pi}*f$, respectively. The corresponding charges are
\begin{equation}
Q_e(\Sigma_2)= \int_{\Sigma_{D-2}} *J_e=\frac{1}{e^2} \int_{\Sigma_{D-2}} *f 
\label{5.5a}
\end{equation}
and 
\begin{equation}
Q_m(\Sigma_{2})= \int_{\Sigma_{2}} *J_m= \frac{1}{2\pi } \int_{\Sigma_{2}} * (*f)= \frac{1}{2\pi } \int_{\Sigma_{2}} f.
\label{5.5b}
\end{equation}
These symmetries will be referred to as
\begin{equation}
U(1)_e^{(1)}\times U(1)_m^{(D-3)}.
\label{5.6}
\end{equation}
We see that there is no magnetic symmetry in $D=2$. In $D=3$, the magnetic symmetry is an ordinary 0-form symmetry and in $D=4$ both electric and magnetic are 1-form symmetries.

The unitary operators that implement the generalized symmetries can be obtained by exponentiation of the charges \eqref{5.5a} and \eqref{5.5b},
\begin{equation}
U_e(\alpha_e,\Sigma_{D-2})=e^{i \alpha_e Q_e(\Sigma_{D-2}) }~~~\text{and}~~~ U_m(\alpha_m, \Sigma_{2})=e^{i \alpha_m Q_m(\Sigma_{2})},
\label{5.6a}
\end{equation}
where $\alpha_e \sim \alpha_e +2\pi$ and $\alpha_m \sim \alpha_m +2\pi$ are the parameters of the transformations.

\begin{figure}
	\centering
	\includegraphics[scale=.7,angle=90]{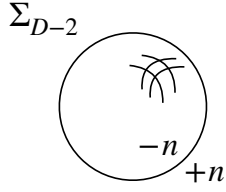}
	\caption{Screening of charges due to the creation of virtual pairs.}
	\label{scre}
\end{figure}
The presence of dynamical charged matter with charge $n >1$ explicitly breaks the $U(1)_e^{(1)}$ symmetry down to $\mathbb{Z}_{n}^{(1)}$, because of screening of the charges due to the creation of virtual pairs. Considering the electric charge $Q_e(\Sigma_{D-2})$ defined in a closed surface $\Sigma_{D-2}$, it is able to detect only charge mod $n$, since it can surround one of the partners of a virtual pair, as illustrated in Fig. \ref{scre}. Therefore, the unitary operator $U_e(\alpha_e,\Sigma_{D-2})$ in \eqref{5.6a} must act as the identity on objects with charges that are multiples of $n$, i.e.,  
\begin{equation}
e^{i \alpha_e Q_e(\Sigma_{D-2}) } = e^{i \alpha_e n k}\equiv 1, ~~~ k\in\mathbb{Z}.
\label{5.6d}
\end{equation}  
This implies that the parameter of the transformation must have the form $\alpha_e = \frac{2 \pi}{n}$ or, equivalently, $e^{i \alpha_e}$ is an element of the $\mathbb{Z}_n$ group.

To further understand the magnetic symmetry it is convenient to parametrize $*f$ in terms of a new gauge field $\tilde{a}$, according to 
\begin{equation}
*f=d\tilde{a},
\label{5.6e}
\end{equation}
where $\tilde{a}$ is a $D-3$ form, namely, 
\begin{equation}
\tilde{a}=\frac{1}{(D-3)!} \tilde{a}_{\mu_1\ldots \mu_{D-2}} dx^{\mu_1}\wedge \cdots dx^{\mu_{D-2}}.
\label{5.6f}
\end{equation}
Relation \eqref{5.6e} implies
\begin{eqnarray}
*f_{\mu_1\ldots\mu_{D-2}}&=&(-1)^{D-1}\partial_{[\mu_{D-2}} \tilde{a}_{\mu_1\ldots \mu_{D-3}]}\nonumber\\
&=&\partial_{[\mu_{1}} \tilde{a}_{\mu_2\ldots \mu_{D-2}]}.
\label{5.6g} 
\end{eqnarray}
Next it is also convenient to express the action in terms of $\tilde{a}$. This should be thought as a change of variables (in the path integral sense) from $a\rightarrow \tilde{a}$. We first change the variables  $a\rightarrow f$, taking into account the Bianchi identity, 
\begin{eqnarray}
S[f,\tilde{a}]&=& \int d^Dx -\frac{1}{4e^2} f_{\mu\nu}f^{\mu\nu} + \frac{1}{2} \tilde{a}_{\mu_1\ldots \mu_{D-3}} \epsilon^{\mu_1\ldots \mu_D}\partial_{\mu_{D-2}} f_{\mu_{D-1}\mu_{D}}
\nonumber\\
&=& \int d^Dx -\frac{1}{4e^2} f_{\mu\nu}f^{\mu\nu} -\frac{1}{2}f_{\mu_{D-1}\mu_{D}}      \epsilon^{\mu_1\ldots \mu_D} \partial_{\mu_{D-2}} \tilde{a}_{\mu_1\ldots \mu_{D-3}} \nonumber\\
&=&  \int d^Dx -\frac{1}{4e^2} f_{\mu\nu}f^{\mu\nu} -\frac12\frac{1}{(D-2)}f_{\mu_{D-1}\mu_{D}}      \epsilon^{\mu_1\ldots \mu_D} \partial_{[\mu_{D-2}} \tilde{a}_{\mu_1\ldots \mu_{D-3}]}\nonumber\\
&=& \int d^Dx -\frac{1}{4e^2} f_{\mu\nu}f^{\mu\nu} -\frac12\frac{(-1)^{D-1}}{(D-2)}f_{\mu_{D-1}\mu_{D}}      \epsilon^{\mu_1\ldots \mu_D}*f_{\mu_1\ldots\mu_{D-2}}(\tilde{a}),
\label{5.6h}
\end{eqnarray}
where $\tilde{a}$ entered initially as a Lagrange multiplier ensuring the Bianchi identity. Now we can integrate out the field $f_{\mu\nu}$ (treated as a basic field) using its equation of motion,
\begin{equation}
f^{\mu_{D-1}\mu_D}= e^2\frac{(-1)^{D}}{(D-2)}   \epsilon^{\mu_1\ldots \mu_D}*f_{\mu_1\ldots\mu_{D-2}}.
\label{5.6i}
\end{equation}
Plugging this back into \eqref{5.6h} and using the identity
\begin{equation}
\epsilon^{\mu_1\ldots\mu_{D-2}\alpha\beta} \epsilon_{\nu_1\ldots\nu_{D-2}\alpha\beta}=2(-1)^{D-1}\left(\delta_{[\nu_1}^{[\mu_1} \delta_{\nu_2}^{\mu_2} \cdots  \delta_{\mu_{D-2]}}^{\mu_{D-2]}}   \right),
\label{5.6j}
\end{equation}
we finally obtain
\begin{equation}
S[\tilde{a}]=\int d^Dx \frac{e^2}{2}  \frac{  (-1)^{D-1} }{(D-2)} *f_{\mu_1\ldots\mu_{D-2}}(\tilde{a}) *f^{\mu_1\ldots\mu_{D-2}}(\tilde{a}).
\label{5.6k}
\end{equation}
This action invariant under the gauge transformations
\begin{equation}
\tilde{a} \rightarrow \tilde{a} + d\lambda.
\end{equation}
In addition to the local gauge-invariant objects obtained from the field strength $*f $, we can construct gauge-invariant extended objects as
\begin{equation}
	T_{q_m}[\Gamma_{D-3}]=\exp \left(i 2\pi q_m \int_{\Gamma_{D-3}} \tilde{a} \right),
	\label{5.6g}
\end{equation}
which are supported on a $(D-3)$-dimensional manifold $\Gamma_{D-3}$, and $q_m$ is the magnetic charge. They are the so-called 't Hooft operators and are the natural candidates to be the charged objects under the magnetic $(D-3)$-form symmetry. We will discuss this explicitly in the four-dimensional case below.

\section{Maxwell in $D=4$}\label{Sec6}

In $D=4$ the higher-form symmetry is $U(1)_e^{(1)}\times U(1)_m^{(1)}$\footnote{As the coupling constant $e^2$ is dimensionless in $D=4$, we set $e^2\equiv1$ for simplicity.}.
The conservation laws in \eqref{5.3} acquire the nicer form
\begin{equation}
\partial_{\mu}f^{\mu\nu}=0~~~\text{and}~~~\partial_{\mu}*f^{\mu\nu}=0.
\label{5.7}
\end{equation}
The charges \eqref{5.5a} and \eqref{5.5b} become
\begin{equation}
Q_e(\Sigma_2)= \int_{\Sigma_2} *J_e= \int_{\Sigma_2} *f 
\label{5.8}
\end{equation}
and 
\begin{equation}
Q_m(\Sigma_2)= \int_{\Sigma_2} *J_m= \frac{1}{2\pi } \int_{\Sigma_2} * (*f)= \frac{1}{2\pi } \int_{\Sigma_2} f,
\label{5.9}
\end{equation}
where $\Sigma_2$ is a closed manifold. The charged operators are Wilson and 't Hooft lines, 
\begin{equation}
W_{q_e}[\mathcal{C}]= \exp \left(i q_e \oint_{\mathcal{C}} a\right) ~~~\text{and}~~~T_{q_m}[\mathcal{C}]=\exp \left(i 2\pi q_m\oint_{\mathcal{C}} \tilde{a}\right).
\label{5.9a}
\end{equation}
With the choice of the $2\pi$ factor in the 't Hooft operator, the magnetic charge $q_m$ is quantized as $q_m \in \mathbb{Z}$, in the same way as the electric charge $q_e$. This can be seen by considering the Wilson line in a closed curve and using the Stokes theorem. In this case, the curve is the boundary of two surfaces, say $X_2$ and $X_2'$, so that 
\begin{equation}
W_{q_e}[\mathcal{C}]= \exp \left(i q_e \oint_{\mathcal{C}} a\right) = \exp \left(i q_e \int_{X_2} f \right)=  \exp \left(i q_e \int_{X_2'} f \right).
\label{5.9b}
\end{equation}
Taking into account the orientation of the surfaces $X_2$ and $X_2'$, the last equality implies
\begin{eqnarray}
1&=&\exp \left(i q_e \int_{\Sigma_2=X_2 \cup X_2'} f \right)\nonumber\\
&=& \exp \left(i 2\pi q_e Q_m(\Sigma_2)\right)=1.
\label{5.9c}
\end{eqnarray}
Thus, we see that the magnetic charges inside $\Sigma_2$ measured by $Q_m(\Sigma_2)$ are integers. With this result in hands, we can find the periodicity of the field $\tilde{a}$ by considering a large gauge transformation. To this, we consider the way it was introduced in the first line of \eqref{5.6h} with $D=4$,
\begin{equation}
S[f,\tilde{a}]=\int d^Dx -\frac{1}{4e^2} f_{\mu\nu}f^{\mu\nu} + \frac{1}{2} \tilde{a}_{\mu} \epsilon^{\mu\nu\rho\sigma}\partial_{\nu}f_{\rho\sigma}.
\end{equation}
Placing this system in a manifold with periodic time $S^1 \times \Omega_3$, with $\partial \Omega_3 =S^2$, a large gauge transformation of $\tilde{a}$ that winds around time direction, $\tilde{a}_0 \rightarrow \tilde{a}_0 +\lambda_0$,  yields
\begin{eqnarray}
\delta S &=& \int_{0}^{L_0} dx^0 \int d^3x \frac{1}{2} \lambda_0 \epsilon^{0ijk} \partial_{i}f_{jk}\nonumber\\
&=& \frac{1}{2} \lambda_0 \int_{0}^{L_0} dx^0 \int_{S^2} dS_i \epsilon^{ijk}f_{jk}\nonumber\\
&=& - \lambda_0 \int_{0}^{L_0} dx^0 \int_{S^2} d\vec{S}\cdot \vec{B}\nonumber\\
&=&-  \lambda_0 L_0 2\pi \mathbb{Z}.
\end{eqnarray}
Invariance of the quantum theory under such a large gauge transformation requires then $e^{i\delta S}=1$, which implies that $\lambda_0$ is of the form
\begin{equation}
\lambda_0 = \frac{n}{L_0},~~~n\in \mathbb{Z}.
\end{equation}
This shows that the large gauge transformations compactify the field $\tilde{a}$ as 
\begin{equation}
\tilde{a}_{\mu} \sim \tilde{a}_{\mu} + \frac{n_{\mu}}{L_{\mu}}.
\end{equation}
Using this large gauge transformation in the 't Hooft line \eqref{5.9a} leads immediately to the quantization of the magnetic charge $q_m \in \mathbb{Z}$.

If we pick up $\Sigma_2=S^2$ immersed in a purely spatial splice, then the above charges are nothing else the electric and magnetic fluxes 
\begin{equation}
Q_e(S^2)=\frac{1}{4} \int_{S^2} dS^{\mu\nu}\epsilon_{\mu\nu\rho\sigma}f^{\rho\sigma} = \int_{S^2} d\vec{S} \cdot \vec{E}
\label{5.10}
\end{equation}
 and
\begin{equation}
Q_m(S^2)=\frac{1}{4 \pi} \int_{S^2} dS^{\mu\nu}f_{\mu\nu} =\frac{1}{2\pi} \int_{S^2} d\vec{S} \cdot \vec{B}.
\label{5.11}
\end{equation}
The charged objects under such charges (objects that links with $S^2$ in $D=4$) are line defects that are extended exclusively along time direction, that is, they correspond just to electric and magnetic charges at rest (in space). Thus, in a particular instant of time, the charges in \eqref{5.10} and \eqref{5.11} can detect the presence of electric and magnetic charges through the flux crossing the surface $S^2$.

\subsection{Canonical Quantization Perspective}

In the canonical quantization perspective we study genuine line operators, i.e., operators extended only along spatial directions. In the absence of charges, 
it is convenient to fix the Coulomb gauge where $a_0=0$ and $\vec{\nabla}\cdot \vec{a}=0$. The canonical momentum is
\begin{equation}
\Pi^i \equiv \frac{\partial \mathcal{L}}{\partial \dot{a}^i}= \dot{a}^i \equiv -E^i.
\label{5.12}
\end{equation}
This implies the following equal-time commutator 
\begin{equation}
[a^i(x),E^j(y)]= - i \delta^{ij}\delta^{(3)}(\vec{x}-\vec{y}).
\label{5.13}
\end{equation}
We still need to implement the Gauss law, $\vec{\nabla}\cdot \vec{E}=0$ (which is equivalent to $\vec{\nabla}\cdot \vec{a}=0$). We see immediately that the above commutation rules are not compatible with the Gauss law. To proceed, we can modify the commutation rule by replacing $\delta^{ij}\rightarrow \delta^{ij}- \frac{\partial_i\partial_j}{\nabla^2}$. Alternatively, we can insist with \eqref{5.13} and impose the Gauss law not at the operator level, but instead as selection rules for physical states, 
\begin{equation}
\vec{\nabla}\cdot \vec{E}\ket{\text{Phys}}=0.
\label{5.14}
\end{equation}
This is more convenient for our discussions.


\subsubsection{Electric Symmetry}

The next step is to construct the unitary operators that act in the Hilbert space. Let us study first the electric symmetry. The corresponding charge operator can be constructed from \eqref{5.8}, simply by picking up $\Sigma_2$ as purely spatial slices, 
\begin{eqnarray}
Q_e(\Sigma_2)&=&\frac{1}{4}\int_{\Sigma_2} (d\Sigma_2)^{\mu\nu}\epsilon_{\mu\nu\rho\sigma}f^{\rho\sigma}\nonumber\\
&=& \int_{\Sigma_2} (d\Sigma_2)_i E^i,
\label{5.15}
\end{eqnarray}
where we have identified $f^{0i}\equiv -E^{i}$ and $(d\Sigma_2)_i =\frac12\epsilon_{ijk}(d\Sigma_2)^{jk}$. We have three charges
\begin{equation}
Q_e^1=\int dx^2 dx^3 E^1,~~~ Q_e^2=\int dx^1 dx^3 E^2,~~~Q_e^3=\int dx^1 dx^2 E^3.
\label{5.16}
\end{equation}
Let us choose one of them, say $Q_e^3$, to study in detail. The corresponding unitary operator can be obtained by exponentiation 
\begin{equation}
U_e(\alpha_e,3)= \exp \left(i \alpha_e Q_e^3 \right),
\label{5.17}
\end{equation}
where $\alpha_e\sim \alpha_e+2\pi$ is the parameter of the transformation. The charged object is the line operator extended along the direction 3:
\begin{equation}
W_{q_e}[\mathcal{C}_3]= \exp \left(-i q_e \int_{\mathcal{C}_3} dy^3 a^{3}\right).
\label{5.18}
\end{equation}
We have to compute 
\begin{equation}
W'_{q_e}[\mathcal{C}_3]=U_e(\alpha_e, 3) W_{q_e}[\mathcal{C}_3] U_e^{\dagger}(\alpha_e, 3).
\label{5.19}
\end{equation}
To this, we recall the BCH theorem in the form
\begin{equation}
e^A e^B = e^{[A,B]} e^B e^A, 
\label{5.20}
\end{equation}
which is valid when $[A,B]$ commutes with both $A$ and $B$. The commutator in our case is
\begin{eqnarray}
[i \alpha_e Q_e^3, -i q_e \int_{\mathcal{C}_3} dy^3 a^{3}]&=& \alpha_e q_e \int dx^1 dx^2 \int_{\mathcal{C}_3}dy^3 [E^3(x^1,x^2,x^3),a^3(y^1,y^2,y^3)]\nonumber\\
&=& i\alpha_e q_e  \int_{\mathcal{C}_3} dy^3 \delta(x^3-y^3),
\label{5.21}
\end{eqnarray}
where we have used the commutation relations in \eqref{5.13}. The integral  $ \int_{\mathcal{C}_3} dy^3 \delta(x^3-y^3)$ is the intersection of the curve $\mathcal{C}_3$ and the plane $\Sigma_2$, which is equal to one when they intersect. Therefore, the transformation of the line operator is
\begin{equation}
U_e(\alpha_e, 3) W_{q_e}[\mathcal{C}_3] U_e^{\dagger}(\alpha_e, 3) = \exp{i\alpha_e q_e \int_{\mathcal{C}_3} dy^3 \delta(x^3-y^3)} W_{q_e}[\mathcal{C}_3].
\label{5.22}
\end{equation}
We notice that the factor in the exponential is essentially the line integral of the Poincaré dual \eqref{3.13}. In fact, in this case, it is given by 
\begin{equation}
\xi_{3}(y)= \int dx^1 dx^2\delta^{(3)}(\vec{x}-\vec{y}).
\label{5.23}
\end{equation}
By integrating it along the curve $\mathcal{C}_3$ gives 
\begin{equation}
\int_{\mathcal{C}_3} dy^3 \xi_{3}(y) = \int_{\mathcal{C}_3} dy^3  \delta(x^3-y^3).
\label{5.24}
\end{equation}
For infinitesimal $\alpha_e$ we see that \eqref{5.22} recovers the transformation given in \eqref{3.12}. The same reasoning goes for the remaining charges $Q_e^1$ and $Q_e^2$.


\subsubsection{Magnetic Symmetry}

Now we will study the magnetic symmetry. To this, it is convenient to write the Maxwell action in terms of the dual field $\tilde{a}$. Putting $D=4$ in \eqref{5.6k}, we have
\begin{equation}
S[\tilde{a}]= \int d^4x -\frac{1}{4} *f_{\mu\nu}*f^{\mu\nu} = \int d^4x -\frac{1}{4} (\partial_{\mu}\tilde{a}_{\nu}-\partial_{\nu}\tilde{a}_{\mu})^2.
\label{5.25}
\end{equation} 
By proceeding with canonical quantization, we first find the momentum
\begin{equation}
\tilde{\Pi}^i \equiv \frac{\partial \mathcal{L}}{\partial \dot{\tilde{a}}^i}=\dot{\tilde{a}}^i \equiv -\tilde{E}^i = - B^i,
	\label{5.26}
\end{equation}
where we have used that $*f_{\mu\nu}=\frac{1}{2}\epsilon_{\mu\nu\rho\sigma}f^{\rho\sigma}$. This implies the following equal-time commutator 
\begin{equation}
	[\tilde{a}^i(x),B^j(y)]= - i \delta^{ij}\delta^{(3)}(\vec{x}-\vec{y}).
	\label{5.27}
\end{equation}

Now we can construct the charges of the magnetic symmetry according to \eqref{5.9}, 
\begin{equation}
Q_m(\Sigma_2)= \frac{1}{2\pi} \int_{\Sigma_2} (d\Sigma_2)_i B^i.
\label{5.28}
\end{equation}
Like in the electric case, we have three charges
\begin{equation}
Q_m^1=\frac{1}{2\pi}\int dx^2 dx^3 B^1,~~~ Q_m^2=\frac{1}{2\pi}\int dx^1 dx^3 B^2,~~~Q_m^3=\frac{1}{2\pi}\int dx^1 dx^2 B^3.
\label{5.29}
\end{equation}
Let us consider the charge $Q_m^3$. The corresponding magnetic unitary operator is
\begin{equation}
	U_m(\alpha_m,3)= \exp \left(i \alpha_m Q_m^3 \right),
	\label{5.30}
\end{equation}
where $\alpha_m\sim \alpha_m+2\pi$ is the parameter of the transformation. The charged object under this symmetry is the line operator
\begin{equation}
	T_{q_m}[\mathcal{C}_3]= \exp \left(-i 2\pi q_m \int_{\mathcal{C}_3} dy^3 \tilde{a}^{3}\right).
	\label{5.30}
\end{equation}
The computation of the transformation law of this 't Hooft operators is essentially the same as that one of the electric case. The result is
\begin{equation}
	U_m(\alpha_m, 3) T_{q_m}[\mathcal{C}_3] U_m^{\dagger}(\alpha_m, 3) = \exp{i \alpha_m q_m \int_{\mathcal{C}_3} dy^3 \delta(x^3-y^3)} T_{q_m}[\mathcal{C}_3].
	\label{5.31}
\end{equation}


\subsection{Algebra of Wilson and 't Hooft Operators}

It is interesting to study the algebra between Wilson and 't Hooft operators in the framework of canonical quantization. To this, we need to express both operators in terms of the same canonical pair. Let us choose to write the 't Hooft operator in terms of electric field. This can be done simply by considering the 't Hooft operator defined in a closed curve in space. Then, using the Stokes theorem we can write
\begin{equation}
T_{q_m}[\mathcal{C}']= \exp \left( - i 2\pi q_m \int_{\mathcal{C}'} d\vec{y}\cdot \vec{\tilde{a}}\right)= \exp \left( - i 2\pi q_m \int_{X_2} \vec{dX_2}\cdot \vec{E}\right),
\label{5.32}
\end{equation}
where $ \vec{d X_2}$ is the oriented integration element on the surface $X_2$. The Wilson operator along a closed or infinitely long purely spatial curve is
\begin{equation}
W_{q_e}[\mathcal{C}] = \exp \left(-i q_e \int_{\mathcal{C}} d\vec{x}\cdot \vec{a}\right).
\label{5.33}
\end{equation}
By using the BCH theorem \eqref{5.20} and the commutation \eqref{5.13}, the algebra of Wilson and 't Hooft operators follows immediately
\begin{equation}
W_{q_e}[\mathcal{C}]T_{q_m}[\mathcal{C}']= \exp \left(i 2\pi q_e q_m \int_{X_2} dX_2^i \int_{\mathcal{C}} dx^i \delta^{(3)}(\vec{x}-\vec{y})  \right) T_{q_m}[\mathcal{C}'] W_{q_e}[\mathcal{C}].
\label{5.34}
\end{equation}
The object $ \int_{X_2} dX_2^i \int_{\mathcal{C}} dx^i \delta^{(3)}(\vec{x}-\vec{y})$ is the intersection of the curve $\mathcal{C}$ and the surface $X_2$, which in turn is equal to the link between the curves,
\begin{equation}
\int_{X_2} dX_2^i \int_{\mathcal{C}} dx^i \delta^{(3)}(\vec{x}-\vec{y})  = \text{Link}(\mathcal{C},\mathcal{C}') = \mathbb{Z}.
\label{5.35}
\end{equation}
Therefore, the relation \eqref{5.34} can be written as
\begin{equation}
	W_{q_e}[\mathcal{C}]T_{q_m}[\mathcal{C}']= e^{i 2\pi q_e q_m \, \text{Link}(\mathcal{C},\mathcal{C}')}T_{q_m}[\mathcal{C}'] W_{q_e}[\mathcal{C}].
	\label{5.36}
\end{equation}
Due to the quantization of the charges $q_e$ and $q_m$, it turns out that the phase factor is equal to one, and the above algebra is commutative.


\subsection{Path Integral Perspective}

\begin{figure}
	\centering
	\includegraphics[scale=.7,angle=90]{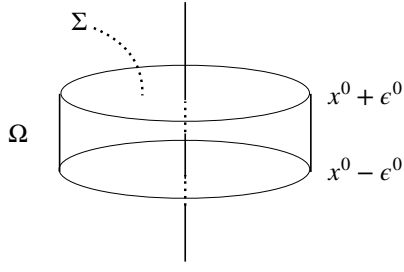}
	\caption{A closed region $\Omega$, in terms of very large spatial surface $\Sigma$. In the limit of $\epsilon^0\rightarrow 0$, we take the surface $\Sigma$ to be infinitely large.}
	\label{region}
\end{figure}

Now we want to derive the transformation of line operators that can be eventually extended along time direction. It is convenient to consider the path integral to compute the correlation function of interest, namely, the correlation function of the unitary operator associated with the charge $Q_e(\Sigma_2)$ acting on a general Wilson line,
\begin{equation}
\langle e^{i \alpha_e Q_e(\Sigma_2)} e^{i q_e \oint_{\mathcal{C}} a} \rangle=  \int \mathcal{D}a\, e^{i \alpha_e Q_e(\Sigma_2) + i q_e \oint_{\mathcal{C}} a + iS},
\label{5.37}
\end{equation}
where $\Sigma_{2}$ is an arbitrary closed $2$-dimensional manifold. The basic idea is that we can absorb $Q_{e}$ into the action through a redefinition of the gauge field \cite{YokokuraSeminar}. The first step is to use the Stokes theorem to write the charge as
\begin{equation}
Q_e(\Sigma_2)= \int_{\Sigma_2} *f = \int_{\Omega} d*f, 
\label{5.38}
\end{equation}
where $\Omega$ is a 3-dimensional volume whose boundary is $\Sigma_2$, i.e., $\partial \Omega=\Sigma_2$. In components, it reads
\begin{equation}
Q_e(\Sigma_2)= \frac{1}{4}\int_{\Omega}  \partial_{\alpha}f^{\mu\nu}\epsilon_{\mu\nu\rho\sigma} dV^{\alpha\rho\sigma}.
\label{5.39}
\end{equation}
We can express the volume element as
\begin{equation}
dV^{\alpha\rho\sigma}= \epsilon^{\alpha\rho\sigma\gamma} n_{\gamma}dV~~~\Leftrightarrow~~~ n_{\gamma}dV= \frac{1}{3!} \epsilon_{\gamma\alpha\rho\sigma}dV^{\alpha\rho\sigma},
\label{5.40}
\end{equation}
where $n_{\gamma}$ is a unit vector normal to the volume $dV^{\alpha\rho\sigma}$. With this, the charge becomes
\begin{equation}
Q_e(\Sigma_2)=\frac{1}{3!}\int_{\Omega} \epsilon_{\nu\rho\sigma\gamma} \partial_{\mu}f^{\mu\nu} dV^{\rho\sigma\gamma}.
\label{5.41}
\end{equation}
Next we turn this expression into an integral over four dimensions with the help of a delta function, 
\begin{eqnarray}
Q_e(\Sigma_2)&=&\int d^4x \partial_{\mu}f^{\mu\nu}(x) J_{\nu}(x;\Omega)\nonumber\\
&=&-\frac{1}{2} \int d^4x f^{\mu\nu}\left[\partial_{\mu}J_{\nu}(x;\Omega)-\partial_{\nu}J_{\mu}(x;\Omega)\right],
\label{5.42}
\end{eqnarray}
where we have identified
\begin{equation}
J_{\nu}(x;\Omega)\equiv \frac{1}{3!}\int_{\Omega} \epsilon_{\nu\rho\sigma\gamma} \delta^{(4)}(x-y)dV^{\rho\sigma\gamma}(y),
\label{5.43}
\end{equation}
which is nonvanishing only when $x \in \Omega$. Notice that this is nothing else the Poincaré dual \eqref{3.8} but with the crucial difference that here the manifold $\Omega$ has a boundary (it is immersed in the four dimensional spacetime). Consequently, $J_{\nu}$ is not a closed form. This is the local version of the global 1-form symmetry transformation, like we have used to derive the Ward identities. If we set $\nu=i$ and choose the region $\Omega$ as depicted in Fig. \ref{region}, the expression \eqref{5.43} reduces to 
\begin{equation}
	J_{i}(x;\Omega)=\int_{\Sigma} \epsilon_{ijk} \delta^{(3)}(\vec{x}-\vec{y})dV^{jk}(y),
	\label{5.43a}
\end{equation}
which is precisely equation \eqref{3.10} with $d=3$.

With the charge expressed as in \eqref{5.42}, the correlation function in \eqref{5.37} reads
\begin{eqnarray}
&&\langle e^{i \alpha_e Q_e(\Sigma_2)} e^{i q_e \oint_{\mathcal{C}} a} \rangle\nonumber\\ 
&=& \int \mathcal{D}a\, \exp\left[ i \int d^4x -\frac{1}{4}f_{\mu\nu}^2- \frac{\alpha_e}{2}f^{\mu\nu}\left[\partial_{\mu}J_{\nu}(\Omega)-\partial_{\nu}J_{\mu}(\Omega)\right] +i q_e \oint_{\mathcal{C}} dx^{\mu}a_{\mu}\right].
\label{5.44}
\end{eqnarray}
Then, under the shift in the gauge field
\begin{equation}
a_{\mu} \rightarrow a_{\mu} -\alpha_e J_{\mu}, 
\label{5.45}
\end{equation}
the expression \eqref{5.44} becomes
\begin{eqnarray}
	&&\langle e^{i \alpha_e Q_e(\Sigma_2)} e^{i q_e \oint_{\mathcal{C}} a} \rangle \\ 
	&=& e^{-i q_e \alpha_e \oint_{\mathcal{C}}dx^{\mu}J_{\mu}(\Omega)} \int \mathcal{D}a\, \exp\left[ i \int d^4x -\frac{1}{4}f_{\mu\nu}^2+ \frac{\alpha_e^2}{2}\left[\partial_{\mu}J_{\nu}(\Omega)-\partial_{\nu}J_{\mu}(\Omega)\right]^2 +i q_e \oint_{\mathcal{C}} dx^{\mu}a_{\mu}\right].\nonumber
	\label{5.46}
\end{eqnarray}
The term proportional to $\int d^4x\left[\partial_{\mu}J_{\nu}(\Omega)-\partial_{\nu}J_{\mu}(\Omega)\right]^2$ is just a local contribution independent on the curve $\mathcal{C}$ and also independent on the dynamical gauge field, so that it can be absorbed into the integration measure. Therefore, we obtain
\begin{equation}
\langle e^{i \alpha_e Q_e(\Sigma_2)} e^{i q_e \oint_{\mathcal{C}} a}\rangle =e^{-i q_e \alpha_e \oint_{\mathcal{C}}dx^{\mu}J_{\mu}(\Omega)} \langle e^{i q_e \oint_{\mathcal{C}} a} \rangle.
\label{5.47}
\end{equation}
Notice that the factor in the exponential outside the correlation function is the intersection of the curve $\mathcal{C}$ and the volume $\Omega$, which in turn is equal to the link 
between the surface $\Sigma_2$ and the curve $\mathcal{C}$,
\begin{equation}
\oint_{\mathcal{C}}dx^{\mu}J_{\mu}(\Omega)= \text{Link}(\Sigma_2,\mathcal{C}).
\label{5.48}
\end{equation}
The relation in \eqref{5.47} is the transformation for a general curve $\mathcal{C}$ in spacetime. We can recover the results of canonical quantization by specifying the curve $\mathcal{C}$ along the spatial directions at fixed time. For example, take $\mathcal{C}$ to be along direction 3. In this case, \eqref{5.48} reduces to
\begin{eqnarray}
\int_{\mathcal{C}_3}dx^{3}J_{3}(\Omega)&=& \int_{\mathcal{C}_3}dx^{3}\int_{\Omega} \epsilon_{3012} \delta^{(4)}(x-y)dV^{012}(y)\nonumber\\
&=& - \int_{\mathcal{C}_3}dx^{3} \delta(x^3-y^3),
\label{5.49}
\end{eqnarray}
where the volume $\Omega$ was taken as $[x^0-\epsilon^0,x^0+\epsilon^0]\times \mathbf{R}_{12}$ and we have used $ \epsilon_{3012}= -\epsilon^{0123}=-1$. Plugging this into \eqref{5.47} reproduces precisely the result obtained in \eqref{5.22}.


\section{Maxwell in $D=3$}\label{Sec7}
				
Free Maxwell theory in $D=3$ possesses a 1-form electric and a 0-form magnetic symmetries. The action is 
\begin{equation}
	S[a]= \int -\frac{1}{2 e^2} f \wedge *f= \int d^3x -\frac{1}{4 e^2} f_{\mu\nu}f^{\mu\nu},
	\label{6.1}
\end{equation}		
where now $e^2$ is dimensionful ($[e^2]=1$) so that we cannot set $e^2=1$. 
		
In this case, the dual field strength $*f$ is a 1-form, 
\begin{equation}
*f=\frac{1}{2 e^2} f_{\mu\nu} *dx^{\mu}\wedge dx^{\nu}= \frac{1}{2e^2} f_{\mu\nu} \epsilon^{\mu\nu}\,_{\rho} dx^{\rho}.
\label{6.2}
\end{equation}		
Relation \eqref{5.5} gives
\begin{equation}
*f_{\alpha} = \frac{1}{2 e^2} f^{\mu\nu} \epsilon_{\mu\nu\alpha}~~~\Leftrightarrow~~~ f_{\mu\nu}=e^2\epsilon_{\mu\nu\rho}*f^{\rho}.
\label{6.3}
\end{equation}		
The electric and magnetic charges are
\begin{equation}
Q_e=\frac{1}{e^2}\int_{\Sigma_1} *f~~~\text{and}~~~ Q_m=\frac{1}{2\pi}\int_{\Sigma_2}f.
\label{6.4} 
\end{equation}
	
We wish to discuss the 't Hooft operators, which in this case turn out to be local operators. In fact, the parametrization \eqref{5.6e} implies that $\tilde{a}$ is a 0-form, 
\begin{equation}
*f=d \tilde{a} ~~~\Rightarrow *f_{\mu}=\partial_{\mu} \tilde{a},
\label{6.5} 
\end{equation}
and the comparison with \eqref{6.3} provides the relation $ \frac{1}{2 e^2} f^{\mu\nu} \epsilon_{\mu\nu\alpha}=\partial_{\alpha}\tilde{a}$. Consequently, the 't Hooft operator \eqref{5.6g} is local
\begin{equation}
T_{q_m}(x)=e^{i 2\pi q_m \tilde{a}(x)}.
\label{6.6}
\end{equation}		
It is usually called a monopole operator, with $\tilde{a}$ referred to as the dual photon. Being a local operator it is conceivable that it can be included in the action, in contrast to the case $D=4$, where it is an extended object and thus cannot enter the action. To appreciate this point, we consider the action in terms of the dual photon given in \eqref{5.6k} for $D=3$, with the inclusion of the monopole operator through a Hermitian combination,
\begin{equation}
S[\tilde{a}]= \int d^3x \,\frac{e^2}{2} (\partial_{\mu} \tilde{a})^2 + \lambda\cos(2\pi q_m \tilde{a}).   
\label{6.7}
\end{equation} 
A simple dimensional analysis shows an interesting feature of the low-energy limit of this model. In the deep IR, where $E\ll e^2$ and $E \ll \lambda^{\frac13}$, which in effect corresponds to $e^2\rightarrow \infty$ and $\lambda \rightarrow \infty$, we see that $\tilde{a}$ is pinned at 0. Small fluctuations around this point are governed by the action
\begin{equation}
	S[\tilde{a}]= \int d^3x \frac{e^2}{2} (\partial_{\mu} \tilde{a})^2 -\frac{1}{2} \lambda (2\pi q_m)^2 \tilde{a}^2,
	\label{6.8}
\end{equation} 
which shows that the theory is gapped! Therefore, the existence of monopole operators changes drastically the low-energy behavior of the theory, opening a gap in the spectrum. A consequence is that the theory is confining \cite{Polyakov:1976fu,Polyakov:1987ez}. 
		
Now we can proceed with canonical quantization. The momentum is
\begin{equation}
\Pi=\frac{\partial \mathcal{L}}{\partial \partial_0\tilde{a}}= e^2 \partial_0\tilde{a}= B,
\label{6.9}
\end{equation}		
implying the equal-time commutation
\begin{equation}
[\tilde{a}(\vec{x}), B(\vec{y}) ]=i \delta^{(2)}(\vec{x}-\vec{y}).
\label{6.10}
\end{equation}
This relation shows that the monopole operator is charged under the magnetic symmetry, whose charge can be written as
\begin{equation}
Q_m=\frac{1}{2\pi} \int d^2x B.
\label{6.11}
\end{equation}
The corresponding unitary operator is
\begin{equation}
U_{m}(\alpha_m)=e^{i \alpha_m Q_m},~~~\alpha_m \sim \alpha_m + 2\pi.
\label{6.12}
\end{equation}
The transformation of the monopole operator is
\begin{equation}
T'_{q_m}(x)= U_{m}(\alpha_m)  T_{q_m}(x) U_{m}^{\dagger}(\alpha_m)= e^{i 2\pi \alpha_m q_m} T_{q_m}(x). 
\label{6.13} 
\end{equation}	
In terms of the field $\tilde{a}$ this corresponds to
\begin{equation}
\tilde{a} \rightarrow \tilde{a} + \alpha_m.
\label{6.14}
\end{equation}
The presence of the monopole operator with charge $q_m$ breaks this symmetry down to a $\mathbb{Z}_{q_m}$, since the parameters of the transformation must be of the form $\alpha_m = \frac{n}{q_m}$, with $n\in \mathbb{Z}$.
		
As a final comment, it is interesting to express the electric symmetry in \eqref{6.4} in terms of the scalar field $\tilde{a}$. It reads
\begin{equation}
Q_e=\frac{1}{e^2}\int_{\Sigma_1} *f = \frac{1}{e^2} \oint \partial_{\mu}\tilde{a} dx^{\mu}.
\label{6.15}
\end{equation}
We see that it measures the winding number of the field $\tilde{a}$ around a closed path. Therefore, the charged operators are vortices of $\tilde{a}$, which are configurations carrying nontrivial
winding (vorticity). This is a manifestation of the particle-vortex duality (see \cite{Tong:2016kpv,Turner:2019wnh} for modern perspectives).

		
\section{Chern-Simons Theory in $D=3$}\label{Sec8}		
		
The Chern-Simons theory in $D=3$ is defined by the action		
\begin{equation}
S_{CS}[a] = \int \frac{k}{4\pi} a da= \int d^3x \frac{k}{4\pi} \epsilon^{\mu\nu\rho} a_{\mu}\partial_{\nu}a_{\rho},
\label{7.1}
\end{equation}		
where $k \in \mathbb{Z}$ is the level of the theory. In general we do not use to think of the gauge field $a$ as the electromagnetic one, but instead as a sort of emergent\footnote{This is a common nomenclature in condensed matter.} $U(1)$ gauge field, referred to as a $U(1)_a$ field. An electromagnetic field, denoted here by capital $A$, can couple to the emergent field and this enables us to associate electric charge to the probe excitations. The corresponding gauge group is represented by $U(1)_A$.

The quantization of the level $k$ follows from large gauge transformations.	We place the system in a manifold $\mathcal{M}=S^1\times S^2$, and define the flux due to the presence of a monopole as in \eqref{5.5b},
\begin{equation}
\frac{1}{2\pi} \int_{S^2} f \in \mathbb{Z}.
\label{7.2}
\end{equation}
We can make large gauge transformations that wind around time direction as in \eqref{5.1b}, with the gauge function $\Lambda= 2\pi\frac{x^0}{L^0}$. This leads to the compactness of the gauge field
\begin{equation}
a_0\sim a_0 + \frac{2\pi}{L^0}.
\label{7.3}
\end{equation}
To see the changing of the action under this transformation it is convenient first to write it as
\begin{equation}
S_{CS}[a]=\int d^3x \frac{k}{4\pi} a_{0} \epsilon^{ij}f_{ij} + \cdots
\label{7.4}
\end{equation}
and then compute the variation,
\begin{equation}
\delta S_{CS}[a]= 2\pi k\int_{0}^{L^0} \frac{dx^0}{L^0} \underbrace{\left( \frac{1}{2\pi} \int_{S^2} d^2x \frac12\epsilon^{ij}f_{ij} \right)}_{n \in \mathbb{Z}}= 2\pi k n, 
\label{7.5}
\end{equation}
which implies that $k \in \mathbb{Z}$ in order $e^{i \delta S_{CS}[a]}$ (quantum theory) to be invariant\footnote{A careful discussion on level quantization can be found in \cite{Witten:2015aoa}.}.

The equations of motion are simply
\begin{equation}
f_{\mu\nu}=0.
\label{7.6}
\end{equation}
This means that there are no local physical degrees of freedom. The only physical (gauge-invariant) degrees of freedom are encoded in the line operators. 

The above equations of motion imply that 2-form current
\begin{equation}
J^{\mu\nu}=\epsilon^{\mu\nu\rho}a_{\rho},
\label{7.7}
\end{equation}
is conserved and consequently the theory has a 1-form symmetry, with charge 
\begin{equation}
Q(\Sigma_1)= \int_{\Sigma_1} *J =  \int_{\Sigma_1} a.
\label{7.8}
\end{equation}
The corresponding unitary operator is 
\begin{equation}
W_{n}[\mathcal{C}] = \exp \left( i n \oint_{\mathcal{C}}a \right),~~~ n \in \mathbb{Z},
\label{7.9}
\end{equation}
which is nothing else a Wilson line.  Notice that invariance under large gauge transformations enforces the parameter $n$ to be an integer, so that the 1-form symmetry is a discrete one. As we shall discuss below, there are further restrictions on the values of $n$.

To further understand the structure of the theory, we proceed with canonical quantization. We write the action as 
\begin{equation}
S= \int d^3x \frac{k}{2\pi} a_2 \partial_0 a_1 +\cdots,
\label{7.10}
\end{equation}
where we have done an integration by parts, and with the dots representing the terms that do not involve time derivative.
The canonical momentum is
\begin{equation}
\Pi_1 = \frac{\partial \mathcal{L}}{\partial \partial_0 a_1}= \frac{k}{2\pi} a_2,
\label{7.11}
\end{equation}
which implies that $a_1$ and $a_2$ form a canonical pair,
\begin{equation}
[a_1(\vec{x}),a_2(\vec{y})]= \frac{2\pi i}{k} \delta^{(2)}(\vec{x}-\vec{y}).
\label{7.12}
\end{equation}
The consequences are interesting. Let us study the algebra of the lines
\begin{equation}
W_1\equiv \exp \left( i \int dx^1 a_1 \right)~~~ \text{and}~~~ W_2 \equiv	\exp \left( i \int dx^2 a_2 \right).
\label{7.13}
\end{equation}
Using the commutation \eqref{7.12}, it follows that 
\begin{equation}
W_1 W_2 = e^{-\frac{2\pi i}{k}} W_2 W_1,
\label{7.14}
\end{equation}
which is precisely a $\mathbb{Z}_k$-symmetry algebra. More generally, we have
\begin{equation}
(W_1)^m (W_2)^n = e^{-\frac{2\pi i m n}{k}}  (W_2)^n (W_1)^m,~~~m, n \in \mathbb{Z}. 
\label{7.15} 
\end{equation}
From this relation we see that $(W_1)^k$ and $(W_2)^k$ behave as the identity (more precisely, a transparent line) in the sense that both commute with  $(W_1)^m $ and $ (W_2)^m$, for all $m$.  In a unified way, the line
\begin{equation}
	W_{k}[\mathcal{C}] = \exp \left( i k \oint_{\mathcal{C}}a \right)
	\label{7.16}
\end{equation}
behaves as the identity from the point of view of the algebra \eqref{7.15} (it does not induce nontrivial holonomy). This implies that the unitary operators associated with the 1-form symmetry are in fact $\mathbb{Z}_k$ operators, thus realizing a discrete $\mathbb{Z}_k^{(1)}$ symmetry.

\begin{figure}
	\centering
	\includegraphics[scale=.7,angle=90]{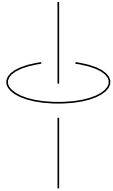}
	\caption{Link between curves as the exchange operation.}
	\label{fig10}
\end{figure}

At first sight, it seems that the lines that induce the same holonomy can be identified, i.e., the lines $n$ and $n+k$ can be identified, but we have to be careful\footnote{A comprehensive discussion can be found in \cite{Seiberg:2016rsg}.}. 
To understand the identification among lines, we need to study their quantum numbers. For this, it is convenient to express the algebra \eqref{7.15} in terms of the link of curves, 
\begin{equation}
W_m[\mathcal{C}] W_n[\mathcal{C}'] = e^{-\frac{2\pi i m n }{k}\text{Link}(\mathcal{C},\mathcal{C}')} W_n[\mathcal{C}'].
\label{7.17} 
\end{equation}
The spin $s$ or, equivalently, the statistics $\nu\equiv 2s$ of a probe particle whose trajectory is represented by a line operator can be derived from this relation. The exchange of particles in the definition of the statistics corresponds to the operation in which one particle goes around the other by an angle of $\pi$ (up to a translation). The statistics is defined in terms of the phase $e^{\pm i \nu \pi}$ that the wave function acquires under exchange. We see that $\nu$ is defined mod 2. 
When we consider a line infinitely extended along time and the another curve encircling the first one at some fixed instant, as in Fig. \ref{fig10}, this corresponds to the operation in which a particle goes around the other by an angle of $2\pi$. Thus, the statistics is half of the value in the exponent in \eqref{7.17}, 
\begin{equation}
\nu = \frac{n^2}{k}~\text{mod}~2~~~\Rightarrow~~~ s=  \frac{n^2}{2k}~\text{mod}~1.
\label{7.18}
\end{equation}
From this it follows that the curves $n$ and $n+k$ can be identified when $k$ is even, 
\begin{equation}
\nu=\frac{(n+k)^2}{k} = \frac{n^2}{k} + 2n +k.
\label{7.19}
\end{equation}
The theory with $k$ even possesses $k$ independent lines. In contrast, when $k$ is odd, the curves $n$ and $n+k$ have distinct statistics (spin) differing mod 1 (mod $\frac12$) and hence cannot be identified, in spite of the fact that they have the same holonomy. In this case, the identification is among the curves $n$ and $n+2k$, implying the existence of $2k$ independent lines. In particular, a line with $n=k$ odd is a fermion with spin $s=\frac{1}{2}$ mod 1. This is important in the description of the fractional quantum Hall effect \cite{Seiberg:2016rsg,Tong:2016kpv}.


\subsection{Coupling to Electromagnetic Field}

Further quantum numbers can be assigned to the line operators. In particular, they are endowed with electric charge. We can see this by coupling the theory with a $U(1)$ electromagnetic field $A$ with the same flux condition as $a$, $\frac{1}{2\pi} \int_{S^2} F \in \mathbb{Z}$. A gauge invariant coupling is of the form $A_{\mu}J^{\mu}$, where $J^{\mu}$ is a conserved current. The only available (1-form) current in the Chern-Simons theory is the topological one,
\begin{equation}
J^{\mu} =\frac{1}{2\pi} \epsilon^{\mu\nu\rho}\partial_{\nu}a_{\rho},
\label{7.20}
\end{equation}
where the normalization is chosen so that the coupling term is invariant under large gauge transformations of both $a$ and $A$.

The expectation value of the Wilson line is equivalent to inserting a term $n a_{\mu} \tilde{J}^{\mu}$ in the action, where the current $\tilde{J}_{\mu}$ of the proble particle can also be parametrized likewise \eqref{7.20}, but with a new gauge field $b$,
\begin{equation}
	\tilde{J}^{\mu} =\frac{1}{2\pi} \epsilon^{\mu\nu\rho}\partial_{\nu}b_{\rho}.
	\label{7.21}
\end{equation}
Thus we consider the action
\begin{equation}
S=\int d^3x \frac{k}{4\pi} \epsilon^{\mu\nu\rho} a_{\mu}\partial_{\nu}a_{\rho} + \frac{1}{2\pi} \epsilon^{\mu\nu\rho} A_{\mu}\partial_{\nu}a_{\rho} +\frac{1}{2\pi} \epsilon^{\mu\nu\rho} a_{\mu}\partial_{\nu}b_{\rho}.
\label{7.22}
\end{equation}
The electric charge of the probe particle can be determined from the coefficient $q_{n}$ of the term $ q_{n} A_{\mu} \tilde{J}^{\mu}$ in the action after the field $a_{\mu}$ is integrated out. The result is
\begin{equation}
\int d^3x \frac{n}{k} A_{\mu} \tilde{J}^{\mu} +\cdots,
\label{7.23}
\end{equation}
showing that the line $W_n$ has electric charge
\begin{equation}
q_{n} = \frac{n}{k},
\label{7.24}
\end{equation}
which is in general fractional. These excitations are identified as the anyons of the  Laughlin phase of the fractional quantum Hall phase \cite{1992IJMPB...6.1711W,Wen_1995}. 


\subsection{Monopole Operators and Bosonization in $D=3$}

We have seen that the 1-form symmetry is a discrete $\mathbb{Z}_k^{(1)}$ symmetry.  Comparison with Maxwell theory, where the $U(1)_e^{(1)}$ is broken down to a $\mathbb{Z}_n^{(1)}$ by the presence of $n$-charged dynamical matter, suggests that there should be $k$-charged dynamical objects in the Chern-Simons theory. They are precisely the monopole operators ('t Hooft operators). To understand this, we recall the definition of a 't Hooft operator in $D=3$  given in \eqref{6.6}, which we repeat here,
\begin{equation}
T(x)=e^{2\pi i \tilde{a}},
\label{7.25}
\end{equation}
where we have set $q_m=1$. The scalar field $\tilde{a}$ enters as a parametrization of the dual field strength, according to $e^2 \partial^{\mu}\tilde{a}=\epsilon^{\mu\nu\rho}\partial_{\nu}a_{\rho}$. With this, we can naively think that the terms $ada$ and $Ada$ in the action \eqref{7.22} represent the couplings $\partial_{\mu}\tilde{a}\, a^{\mu}$ and $\partial_{\mu}\tilde{a}\, A^{\mu}$ of the monopole operator with both gauge fields $a$ and $A$. This implies that the monopole operator is charged under both $U(1)_a$ and $U(1)_A$. The respective charges can be determined by using the equations of motion. 

To do this, we first introduce in the action \eqref{7.22} a Maxwell term for the $a$ field as $-\frac{1}{4g^2}f_{\mu\nu}^2$, with the coupling constant $g^2$ (we reserve $e$ for electromagnetic coupling). The equations of motion are
\begin{equation}
\frac{1}{g^2}\partial_{\mu}f^{\mu\nu}+ \frac{k}{2\pi}\epsilon^{\nu\alpha\beta}\partial_{\alpha}a_{\beta}+ \tilde{J}^{\nu}=0.
\label{7.26}
\end{equation}
Picking up the component $\nu=0$, we get
\begin{equation}
\frac{1}{g^2} \vec{\nabla}\cdot \vec{e} +  \frac{k}{2 \pi} \epsilon^{ij}\partial_i a_j + \tilde{J}^0=0.
\label{7.27} 
\end{equation} 
This relation shows that the Chern-Simons term provides charge $k$ under $U(1)_a$ to the monopole operator. 

Next we introduce a Maxwell term for the electromagnetic field and proceed similarly. The equations of motion are
\begin{equation}
	\frac{1}{e^2}\partial_{\mu}F_A^{\mu\nu}+ \frac{1}{2\pi}\epsilon^{\nu\alpha\beta}\partial_{\alpha}a_{\beta}=0,
	\label{7.28}
\end{equation}
which show that the monopole operator has charge 1 under $U(1)_A$.

We can use the above results to unveil an interesting perspective that leads to a special type of duality in $D=3$, namely, a bosonization duality. Suppose we turn the probe particles associated with $\tilde{J}$ in \eqref{7.22} into dynamical ones, and assume that they are represented by a complex scalar field with charge $1$ under $U(1)_a$. The corresponding Lagrangian reads
\begin{equation}
\mathcal{L}=|(\partial_{\mu}-i a_{\mu}) \phi|^2 + \frac{k}{4\pi} \epsilon^{\mu\nu\rho} a_{\mu}\partial_{\nu}a_{\rho} + \frac{1}{2\pi} \epsilon^{\mu\nu\rho} A_{\mu}\partial_{\nu}a_{\rho} +\cdots.
\label{7.28a}
\end{equation}
We can construct the following object,
\begin{equation}
 \phi^{\dagger}\,T(x),
\label{7.29} 
\end{equation} 
which is the monopole operator dressed by the complex field. It represents a bound-state of a $\phi$-particle and a monopole.  Notice that this object is charged under $U(1)_A$ with charge 1, and uncharged under $U(1)_a$ for $k=1$. Moving one bound-state around another one induces an Aharonov-Bohm phase 
\begin{equation}
e^{i \text{(charge under $U(1)_A$)} \times \text{(flux unit of $B_A$)} }= e^{i  (1) \times (2\pi)},
\end{equation} 
where $B_A=\epsilon^{ij}\partial_iA_j$, from which we can extract the statistics $\nu =1$. Therefore, the composite object $ \phi^{\dagger}\,T(x)$ possesses precisely the quantum numbers of an electromagnetically charged fermion, which suggests the following duality relation  
\begin{equation}
|(\partial_{\mu}-i a_{\mu}) \phi|^2 + \frac{k}{4\pi} \epsilon^{\mu\nu\rho} a_{\mu}\partial_{\nu}a_{\rho} + \frac{1}{2\pi} \epsilon^{\mu\nu\rho} A_{\mu}\partial_{\nu}a_{\rho} +\cdots ~~\Leftrightarrow~~ \bar{\psi} (\slashed{\partial} - i \slashed{A})\psi +\cdots.
\end{equation} 
With the proper refinements (relegated to the dots), this is the bosonization duality in $D=3$, and it figures as the heart of the so-called web of dualities \cite{Karch:2016sxi,Murugan:2016zal,Seiberg:2016gmd}.


\subsection{Bosonization in $D=2$}

The Chern-Simons action \eqref{7.1} in gauge-invariant up to boundary terms. Therefore, in a manifold with boundary, the gauge freedom cannot be used at the boundaries to remove degrees of freedom so that it is expected physical modes living at the boundaries.  This fact can be used to understand bosonization in $D=2$\footnote{See \cite{vonDelft:1998pk,Senechal:1999us} for detailed discussions on bosonization in $D=2$.}. 

We consider the manifold defined in the semi-plane $x^2 \in (-\infty, 0]$, so that there is a physical boundary at $x^2=0$, and parametrize the solution of \eqref{7.6} in terms of a scalar field as $a_{\mu}=\partial_{\mu} \phi$, i.e., a pure gauge configuration in the bulk. Next, we take the Wilson line along an open curve coming from infinity and ending at the boundary, and use this parametrization to write it as
\begin{eqnarray}
W_n[x^0,x^1] &=& \exp \left(i n \int_{-\infty}^{0} dx^2 a_2\right) \nonumber\\
&=& \exp \left(i n \phi(x^0,x^1) \right).
\label{8.1}
\end{eqnarray}
This edge operator is usually referred to as a vertex operator in CFT language \cite{DiFrancesco:1997nk}.

In the commutator \eqref{7.12}, we set $x^2=0$ and integrate both sides in $y^2$ in the interval $x^2 \in (-\infty, 0]$, which gives
\begin{equation}
[\partial_1\phi (x^1),\phi(y^1)]= \frac{2\pi i}{k} \delta^{(1)}(x^1-y^1).
\label{8.2}
\end{equation}
This implies 
\begin{equation}
[\phi (x^1),\phi(y^1)]=\frac{\pi i}{k} \text{sign}(x^1-y^1),
\label{8.2a}
\end{equation}
which is the commutation relation for a chiral boson \cite{Floreanini:1987as} (more recent discussions can be found in \cite{Frishman:2010zz}).
		
Comparing \eqref{8.2} with the commutation rule for a chiral system, namely,
\begin{equation}
[\phi(x^1), \Pi(y^1)]= \frac{i}{2} \delta^{(1)}(x^1-y^1),
\label{8.3}
\end{equation}
we see that the momentum is identified as $\Pi=-\frac{k}{4\pi } \partial_1 \phi$.

Now we look for a Lagrangian that gives this relation. It must contain
\begin{equation}
\mathcal{L} = -\frac{k}{4\pi} \partial_0\phi \partial_1\phi +\cdots.
\label{8.4}
\end{equation}
If we do not include any other term in the dots, the theory has no propagating degree of freedom. A sensible choice is
\begin{equation}
\mathcal{L} = -\frac{k}{4\pi}\left( \partial_0\phi \partial_1\phi + v (\partial_{1}\phi)^2 \right), 
\label{8.5}
\end{equation}
where we have included a new parameter $v$ that represents the velocity of propagation in the boundary. The equation of motion is
\begin{equation}
(\partial_0 + v \partial_1) \partial_1\phi=0,
\label{8.5a}
\end{equation}
which shows that $\phi$ is indeed a chiral field, i.e., one-way propagating.

The corresponding Hamiltonian is 
\begin{equation}
H=\int dx^1 \frac{k}{4\pi} v (\partial_1\phi)^2.
\label{8.6}
\end{equation}		
It is positive definite only if $k v >0$. Thus, the sign of the Chern-Simons level $k$ determines the sign of the velocity $v$.
		
Let us study the statistics of the vertex operators in \eqref{8.1}.	According to the commutation in \eqref{8.2a}, 
\begin{equation}
e^{i n \phi(x^1)} e^{i m \phi(y^1)}= \exp\left( - \frac{\pi i n m}{k} \text{sign}(x^1-y^1) \right) e^{i m \phi(y^1)} e^{i n \phi(x^1)}.
\label{8.7}
\end{equation}	
This means that a vertex operator $e^{i n \phi(x)}$ has statistics $e^{- \pi i \frac{n^2}{k}}$. In particular, the vertex operator with $n=k$ is a fermion if $k$ is odd ($e^{-\pi i k}=-1$ for odd $k$).

In the simplest case $k=1$ (at the compactification radius equal to one), an explicit version of the fermionized theory can be achieved. In this case, we can write
\begin{equation}
\psi \equiv c \, e^{i \phi},
\label{8.8}
\end{equation}
where $c$ is some normalization constant. Dimensional analysis shows that $c$ must have dimension $[c]=\frac12$ in mass units. Thus we write is 
\begin{equation}
	\psi \equiv \frac{1}{\sqrt{2\pi a}} \, e^{i \phi},
	\label{8.8a}
\end{equation}
where $a$ is a short-distance cutoff. Let us compute 
\begin{eqnarray}
\psi^{\dagger}(x') \psi(x) &=& \frac{1}{2\pi a} e^{-i  \phi(x')} e^{i  \phi(x)}\nonumber\\
&=& \frac{1}{2\pi a}  e^{- i (\phi(x') -\phi(x) )} e^{\frac{\pi i  }{2}\text{sign}(x'-x )}.
\label{8.9}
\end{eqnarray}
We are interested in the limit $x'\rightarrow x$, but this should be done carefully. We then set $x'=x+a$, 
\begin{eqnarray}
	\psi^{\dagger}(x') \psi(x) &=&  \frac{1}{2\pi a} e^{\frac{\pi i  }{2} \text{sign}(a)}  \left(1 -i a \partial_x\phi - \frac{i  a^2}{2}\partial_x^2\phi -\frac{ a^2}{2} (\partial_x\phi)^2+\cdots \right).
	\label{8.10}
\end{eqnarray}
Next we apply a derivative with respect to $x$ in both sides, taking into account that in the right hand side $a$ is now dependent on $x$, so that $\partial_x a =-1$. The only nontrivial contribution comes when the derivative acts on the factor $\frac{1}{2\pi a}$, resulting in 
\begin{equation}
\psi^{\dagger}(x') \partial_x\psi(x) =  - i  \frac{1}{4\pi }(\partial_x\phi)^2+\cdots.
\label{8.11}
\end{equation}
The terms in the dots can be neglected since they are constants (divergent), total derivatives, or even vanish in the limit $a\rightarrow 0$. With this, we have obtained essentially the bosonization formula for the Hamiltonian \eqref{8.6}, which in terms of fermionic field becomes
\begin{equation}
	H=- \int dx \,i \, v\, \psi^{\dagger}(x) \partial_x\psi(x).
	\label{8.12}
\end{equation}		
The corresponding Lagrangian is
\begin{equation}
\mathcal{L}= i \psi^{\dagger} (\partial_0+v\partial_x) \psi,
\label{8.13}
\end{equation}
which is the Lagrangian of a chiral free fermion \cite{Floreanini:1987as,Frishman:2010zz}.


\subsection{'t Hooft Anomaly and Ground State Degeneracy in the Torus}

Let us go back to the algebra \eqref{7.15} to discuss some further implications. It is a projective representation of the Abelian algebra, in the sense that it fails to reduce to a commutator because of a phase. A projective representation signals the existence of an 't Hooft anomaly, i.e., an obstruction to gauging, and has consequences on the spectrum of the theory. In fact, an 't Hooft anomaly famously leads to the  so-called matching conditions \cite{tHooft:1979rat}. Specifically, an 't Hooft anomaly in the UV cannot be matched by a trivial gapped theory in the IR, so that the IR must be nontrivial. 

Let us write generically the projective representation of an algebra involving two unitary operators as
\begin{equation}
U_1 U_2 = e^{i \alpha } U_2 U_1. 
\label{9.1}
\end{equation}
Next consider the following two states, 
\begin{equation}
\ket{\psi'}= U_1 U_2 \ket{\psi}~~~\text{and} ~~~ \ket{\psi''}= U_2 U_1 \ket{\psi},
\label{9.2} 
\end{equation}
constructed out the same original physical state $\ket{\psi}$. According to the algebra \eqref{9.1}, we see that
\begin{eqnarray}
\ket{\psi'} = e^{i\alpha}  \ket{\psi''}.
\label{9.3}
\end{eqnarray}
As the two states differ only by a phase, they belong to the same ray and are actually identified as the same physical state. 

Now suppose that we gauge the global symmetries associated with $U_1$ and $U_2$. This means that physical states must be invariant under action of such operators, i.e., $U_1\ket{\psi}=\ket{\psi} $ and  $U_2\ket{\psi}=\ket{\psi}$. Thus, the algebra in \eqref{9.1} implies 
\begin{equation}
(1-e^{i\alpha}) \ket{\psi}=0. 
\label{9.4}
\end{equation}
As in general $e^{i\alpha} \neq 1$, it follows that there are no notrivial physical states left behind in the theory, $\ket{\psi}=0$. In particular, the partition function vanishes.

Now we turn to the Chern-Simons theory. In this case, the gauging of the $\mathbb{Z}^{(1)}_k$ symmetry can be done in the following way. We introduce in the action the coupling $c_{\mu\nu} J^{\mu\nu}$, where $J^{\mu\nu}$ is the 2-form current \eqref{7.7} and $c_{\mu\nu}$ is a flat gauge connection for the $\mathbb{Z}^{(1)}_k$-symmetry\footnote{In gauging a {\it continuous} symmetry, there is also curvature.}, and sum over $c_{\mu\nu}$ in the partition function.  Being a flat connection, we can write it as
\begin{equation}
c_{\mu\nu}= \frac{1}{4\pi}\partial_{[\mu}c_{\nu]}. 
\label{9.5}
\end{equation}
Let us treat initially the gauge field as a background one. In this case, the partition function reads
\begin{equation}
Z[c]=\int \mathcal{D}a \exp \left[i \int d^3x  \left(\frac{k}{4\pi}\epsilon^{\mu\nu\rho}a_{\mu}\partial_{\nu}a_{\rho}+\frac{1}{2\pi}\epsilon^{\mu\nu\rho}a_{\mu}\partial_{\nu}c_{\rho}\right)\right].
\label{9.6}
\end{equation}
Identifying the current $J^{\mu}\equiv \frac{1}{2\pi}\epsilon^{\mu\nu\rho}\partial_{\nu}c_{\rho}$, we can express the term $\int d^3x a_{\mu} J^{\mu}$ in the action in terms of a line integral $\oint a$, as discussed in the derivation of \eqref{evwl}. In this way,
\begin{eqnarray}
	Z[c]&=&\int \mathcal{D}a \exp \left[i \int d^3x \frac{k}{4\pi}\epsilon^{\mu\nu\rho}a_{\mu}\partial_{\nu}a_{\rho}+i n \oint_{\mathcal{C}} a \right]\nonumber\\
	&=& \langle W_n[\mathcal{C}]\rangle.
\end{eqnarray} 
Now we convert to the Hilbert space perspective. To this, we choose the line $\mathcal{C}$ to be purely spatial, say infinitely extended along direction 1\footnote{According to the notation of \eqref{7.13}, $W_n[1]=(W_1)^n$.}, and rotate to the Euclidean periodic time $\tau$ (imaginary time formalism),
\begin{equation}
Z[n]= \int  \mathcal{D}a\, W_n[1] \, e^{-\int_0^{\beta}d \tau d^2x \mathcal{L} }.
\end{equation}
The information of the background gauge field remaining in this expression is encoded in $n$. Therefore, gauging the theory means sum over all possible operators $W_n[1]$, namely,  
\begin{equation}
Z_{gauged}=\sum_{n=0}^{k-1}\int  \mathcal{D}a\, W_n[1] \, e^{-\int_0^{\beta}d \tau d^2x \mathcal{L} }.
\end{equation}
This expression, in turn, can be written as
\begin{equation}
Z_{gauged}= \text{Tr}\left( e^{-\beta H}\sum_{n=0}^{k-1} W_n[1]  \right),
\end{equation}
where $H$ is the Chern-Simons Hamiltonian, which in turn vanishes, but this is innocuous in the present derivation.
Then, picking up a basis that diagonalizes the unitary operators $W_n[1]$, that is, $\ket{m}\equiv (W_2)^m \ket{0}$ (see the discussion after \eqref{9.14}),
the partition function is written as
\begin{equation}
Z_{gauged}= \sum_{m=0}^{k-1} \bra{m} e^{-\beta H}  (1+ e^{-\frac{2\pi i m}{k} } + e^{-\frac{2\pi i 2 m}{k} }+\cdots +e^{-\frac{2\pi i (k-1)m}{k} } )  \ket{m},
\end{equation}
which vanishes for any $k>1$. 

The above discussion shows that there is an obstruction to gauging the  $\mathbb{Z}^{(1)}_k$-symmetry due to a 't Hooft anomaly. Now we discuss the consequences of it in the spectrum. We wish to compute the ground state degeneracy of the Chern-Simons theory in a spatial torus, $\mathcal{M}= \mathbf{R} \times T^2$.

Let us consider a torus as a rectangle of sizes $L_1$ and $L_2$, with the identifications $x^1 \sim x^1 +L_1$ and $x^2 \sim x^2 +L_2$. In this case, we can decompose the gauge field in Fourier modes as
\begin{equation}
a_{\mu}(x^0,\vec{x})= \sum_{\vec{k}} e^{i \vec{k}\cdot \vec{x}} a_{\mu}(x^0,\vec{k}),
\label{9.9}
\end{equation}
where $k^1 = 2\pi \frac{n^1}{L_1}$ and $k^2 = 2\pi \frac{n^2}{L_2}$. To study the ground state degeneracy, we need to consider only the zero modes,
\begin{equation}
a_{\mu}(x^0,\vec{x})=a_{\mu}(x^0,\vec{k}=0)+ \cdots,
\label{9.10}
\end{equation}
since the nonzero modes are separated by gaps of the order $\frac{1}{L_1}, \frac{1}{L_2}$ \footnote{In the strict case of pure Chern-Simons theory the nonzero modes are in fact gauge modes since the vacuum solutions of the equations of motion are of the form $a_i = \partial_i\phi + \bar{a}_i(t)/L^i$.}. Then, the Chern-Simons theory reduces to a simple quantum mechanical system. 
We redefine the zero modes to absorb the lengths of the torus $\bar{a}_i \equiv a_i (x^0,\vec{k}=0) L^i$ (no sum). The line operators \eqref{7.13} around the two holonomies of the torus reduce to
\begin{equation}
W_1= e^{i \bar{a}_1} ~~~\text{and}~~~W_2= e^{i \bar{a}_2},
\label{9.11}
\end{equation}
with the algebra  given by \eqref{7.14},
\begin{equation}
	W_1 W_2 = e^{-\frac{2\pi i}{k}} W_2 W_1,
	\label{9.12}
\end{equation}
representing the algebra of the ground state. Thus, the ground state degeneracy follows from the size of the representation implied by \eqref{9.12}. Suppose we choose to diagonalize the operator $W_1$, 
\begin{equation}
W_1 \ket{0}  = e^{i\lambda} \ket{0}.
\label{9.13} 
\end{equation}
Then, the operator $W_2$ plays the role of creation operators, 
\begin{equation}
\ket{p} \equiv (W_2)^p \ket{0}.
\label{9.14}
\end{equation}
The algebra in \eqref{9.12} leads to
\begin{equation}
W_1 (W_2)^p \ket{0} = e^{-\frac{2\pi i p}{k}} e^{i\lambda}   (W_2)^p \ket{0},
\label{9.15}
\end{equation}
which shows that there are $k$ distinct states $\ket{p}$, labeled by $p=0,1,\ldots,k-1$. Using \eqref{9.12} we can see immediately that they are linearly independent, 
\begin{eqnarray}
\langle p | p'\rangle &=& \bra{0} (W_2)^{-p } (W_2)^{p'} \ket{0}\nonumber\\
&=& \bra{0} (W_2)^{-p } W_1^{-1}W_1 (W_2)^{p'} \ket{0} \nonumber\\
&=& e^{\frac{2\pi i (p-p')}{k}}  \bra{0} (W_2)^{-p } (W_2)^{p'} \ket{0},
\label{9.16}
\end{eqnarray}
i.e., the scalar product vanishes unless $p=p'$. Therefore, the ground state degeneracy in the tours is $k$. This result can be easily generalized to the case of a $g$-torus, i.e., a surface with genus $g$. It is equivalent to the composition of a number $g$ of torus, so that  the ground state degeneracy is simply
\begin{equation}
\underbrace{k \times \cdots \times k}_{g~ \text{times}} = k^g.
\label{9.17}
\end{equation}


\section{1-Form Symmetry in Non-Abelian Gauge Theories}\label{Sec9}

Our main concern here is the case of four-dimensional non-Abelian gauge theories with gauge group $SU(N)$, governed by the Yang-Mills action 
\begin{equation}
	S_{YM}[a]= \int -\frac{1}{e^2} \text{Tr}( f \wedge *f )= \int d^4x -\frac{1}{2 e^2} \text{Tr}( f_{\mu\nu}f^{\mu\nu}),
	\label{10.1}
\end{equation}
where the non-Abelian field strength is defined as
\begin{equation}
	f_{\mu\nu} = \partial_{\mu} a_{\nu} - \partial_{\nu} a_{\mu} - i [a_{\mu},a_{\nu}].
	\label{10.2}
\end{equation}
The gauge field $a_{\mu}$ is algebra-valued, 
\begin{equation}
a_{\mu}= T^a a_{\mu}^a,
\label{10.3}
\end{equation}
with the Hermitian generators $T^a$ satisfying
\begin{equation}
[T^a,T^b]= i \mathrm{f}^{abc}T^c~~~\text{and}~~~ \text{Tr}(T^a T^b) =\frac{1}{2}\delta^{ab}.
\label{10.4}
\end{equation}

Under a gauge transformation
\begin{equation}
a_{\mu} \rightarrow a_{\mu}'= U(x) a_{\mu} U^{\dagger}(x) + i  U(x) \partial_{\mu}U^{\dagger}(x),
\label{10.5}
\end{equation}
with $U\in SU(N)$, the field strength transforms as
\begin{equation}
f_{\mu\nu}\rightarrow f'_{\mu\nu}= U(x) f_{\mu\nu} U^{\dagger}(x),
\label{10.6}
\end{equation}
which leaves the action \eqref{10.1} invariant. 

The equations of motion of the action \eqref{10.1} are
\begin{equation}
D_{\mu}f^{\mu\nu a}= \partial_{\mu}f^{\mu\nu a}+\mathrm{f}^{abc} a_{\mu}^b f^{\mu\nu c}=0.
\label{10.6a}
\end{equation}
In terms of matrices, this can be written as 
\begin{equation}
D_{\mu} f^{\mu\nu}= \partial_{\mu}f^{\mu\nu} - i [a_{\mu}, f^{\mu\nu}]=0.
\label{10.6b} 
\end{equation}
More generally, the covariant derivative of any object $\phi=\phi^aT^a$ belonging to the adjoint representation is
\begin{equation}
D_{\mu} \phi= \partial_{\mu}\phi- i [a_{\mu}, \phi].
\label{10.6c}
\end{equation}

As in the Abelian case, here we also have the Bianchi identity. In terms of the Hodge dual 
\begin{equation}
*f^{\mu\nu}=\frac{1}{2} \epsilon^{\mu\nu\rho\sigma}f_{\rho\sigma},
\label{10.6d}
\end{equation}
it is given by
\begin{equation}
D_{\mu}*f^{\mu\nu}=0.
\label{10.6e}
\end{equation}


\subsection{Wilson Lines}

With this basic setup, we are ready to investigate the Wilson line operators in the non-Abelian case \cite{Srednicki:2007qs}. It requires a slight generalization compared with the Abelian case. To see this, we consider first the Wilson operator along an infinitesimal path of length $\epsilon$,
\begin{equation}
W(x+\epsilon,x)= e^{ i \epsilon_{\mu} a^{\mu}(x)}= \openone +  i  \epsilon_{\mu} a^{\mu}(x)+\cdots.
\label{10.7}
\end{equation}
Upon a gauge transformation \eqref{10.5}, this object transforms as
\begin{eqnarray}
W(x+\epsilon,x)\rightarrow W'(x+\epsilon,x)&=& \openone + i \epsilon^{\mu} U(x) a_{\mu} U^{\dagger}(x) - \epsilon^{\mu} U(x) \partial_{\mu} U^{\dagger}(x)+\cdots\nonumber\\
&=&  \openone + \epsilon^{\mu} \partial_{\mu} U(x)  U^{\dagger}(x) +  i \epsilon^{\mu} U(x) a_{\mu} U^{\dagger}(x) +\cdots\nonumber\\
&=& [(\openone+\epsilon^{\mu}\partial_{\mu}) U(x)] U^{\dagger}(x) +  i \epsilon^{\mu} U(x) a_{\mu} U^{\dagger}(x) +\cdots\nonumber\\
&=& U(x+\epsilon) (\openone +  i  \epsilon_{\mu} a^{\mu}(x))U^{\dagger}(x)+ \cdots.
\label{10.8}
\end{eqnarray}
Therefore, we see that 
\begin{equation}
W(x+\epsilon,x)\rightarrow W'(x+\epsilon,x) = U(x+\epsilon)W(x+\epsilon,x)U^{\dagger}(x).
\label{10.9}
\end{equation}
Note that $W^{\dagger}(x+\epsilon,x)=W(x-\epsilon,x)$. On the other hand, $W(x-\epsilon,x)$ is equal $W(x,x+\epsilon)$ up to second order terms in $\epsilon$, so that we can write
\begin{equation}
W^{\dagger}(x+\epsilon,x) = W(x,x+\epsilon).
\label{10.10}
\end{equation}
This implies that the Hermitian conjugate reverses the orientation of the path. 

Now we can consider the Wilson line operator along a finite path through the composition of many infinitesimal connected pieces. Consider two points $x$ and $y$, with $y$ reached after $n$ infinitesimal displacements from $x$, i.e., $y=x+\epsilon_1+\epsilon_2+\cdots+\epsilon_n$. We define the string of infinitesimal Wilson operators as
\begin{equation}
W_P(y,x)\equiv W(y,y-\epsilon_n) W(y-\epsilon_n,y-\epsilon_n-\epsilon_{n-1})\ldots W(x+\epsilon_1+\epsilon_2,x+\epsilon_1)W(x+\epsilon_1,x),
\label{10.11}
\end{equation} 
where $P$ stands for the path ordering specified by the right hand side. The path ordering is important to ensure a nice transformation property of $W_P(y,x)$. Under the gauge transformation \eqref{10.5}, $W_P(y,x)$ transforms as
\begin{equation}
W_P(y,x)\rightarrow W'_P(y,x) = U(y) W_P(y,x) U^{\dagger}(x).
\label{10.12}
\end{equation}
In view of \eqref{10.10}, it follows that 
\begin{equation}
W_P^{\dagger}(y,x) = W_{-P}(x,y),
\label{10.13}
\end{equation}
where $-P$ means the reverse of the path $P$. 

With the above results, we can construct a gauge invariant Wilson line operator. We just consider the continuum limit of a closed oriented path and then take the trace, 
\begin{equation}
W[\mathcal{C}] = \text{Tr} P e^{i \oint_{\mathcal{C}}a},
\label{10.14}
\end{equation}
where $P$ inside the trace stands for path ordering.
It is important to emphasize that implicitly there is an assumed representation in this expression, because the generators are given in some representation and the trace must be taken accordingly. 
We shall discuss more about this below.


\subsection{1-Form Center Symmetry}

\subsubsection{Fundamental Representation}

The center subgroup of $SU(N)$ can be constructed in a simple way \cite{nair2005quantum}. Let us consider the fundamental representation and we choose one of the generators, say the last one, to be diagonal. It can be written as
\begin{eqnarray}
T^{N^2-1}=\sqrt{\frac{N}{2(N-1)}} \,\text{diag} \left(\frac{1}{N},\frac{1}{N},\ldots,\frac{1}{N}, -1+\frac{1}{N} \right),
\label{10.34}
\end{eqnarray}
with the diagonal elements ensuring that its trace vanishes. 

Then consider the group element
\begin{equation}
U = e^{i \,\theta\, T^{N^2-1}},
\label{10.35}
\end{equation}
and choose the parameter as $\theta \equiv 2\pi \sqrt{\frac{2(N-1)}{N}}$. It is immediate to see that this group element is proportional to the identity, being an element of the $\mathbb{Z}_N$ group,  
\begin{equation}
U = e^{2\pi i t^{N^2-1}} =e^{\frac{2\pi i}{N}}\openone,
\label{10.36}
\end{equation}
where we have defined the rescaled generator
\begin{equation}
t^{N^2-1}\equiv  \sqrt{\frac{2(N-1)}{N}}T^{N^2-1} =\text{diag} \left(\frac{1}{N},\frac{1}{N},\ldots,\frac{1}{N}, -1+\frac{1}{N} \right).
\end{equation}
Note that the matrix $U$ in \eqref{10.36} has unit determinant, as it should be.

To unveil the global center symmetry of the $SU(N)$ gauge theory it is convenient to consider a manifold with periodic time $S^1$, which amounts to the identification $x^0 \sim x^0+L^0$. In this case, a large gauge transformation that winds around the time direction reads
\begin{equation}
U(x^0) = e^{2\pi N i \frac{x^0}{L^0} t^{N^2-1}}.
\label{10.37}
\end{equation}
This is well-defined under $x^0 \rightarrow x^0+L^0$ in the sense that 
\begin{equation}
U(x^0+L^0)=U(x^0), 
\label{10.37a}
\end{equation}
which follows because $e^{2\pi N i t^{N^2-1}}=\openone$. Therefore, it is a true (large) gauge transformation.

Now we can consider a modification of \eqref{10.37a}, involving the twist by an element of the center $\mathbb{Z}_N$ of $SU(N)$, i.e., 
\begin{equation}
\tilde{U}(x^0+L^0) = \tilde{U}(x^0) h, ~~~h \in \mathbb{Z}_N.
\label{10.38}
\end{equation}
An explicit matrix-valued function $ \tilde{U}(x^0)$ satisfying this condition is
\begin{equation}
	\tilde{U}(x^0) = e^{2\pi k i \frac{x^0}{L^0} t^{N^2-1}}, ~~~k=0,1,\ldots, N-1.
	\label{10.39}
\end{equation}
It is immediate to see that it satisfies 
\begin{equation}
\tilde{U}(x^0+L^0) = \tilde{U}(x^0)e^{\frac{2\pi i k}{N}}.
\label{10.40}
\end{equation}
Notice that a transformation with $\tilde{U}$,
\begin{eqnarray}
a_{0} \rightarrow a_{0}'&=& \tilde{U}(x^0) a_{0} \tilde{U}^{\dagger}(x^0) + i  \tilde{U}(x^0) \partial_{0}\tilde{U}^{\dagger}(x^0)\nonumber\\
&=&  \tilde{U}(x^0) a_{0} \tilde{U}^{\dagger}(x^0) + \frac{2\pi k }{L^0} t^{N^2-1},
\label{10.41}
\end{eqnarray}
does not correspond to a gauge transformation, since it cannot be removed by any kind of gauge transformation (not even a large gauge one). Using \eqref{10.6}, it is immediate to see that it is a true global symmetry of the Yang-Mills action. As we shall see, it is a 1-form symmetry since it acts on the Wilson line operators.

Consider a Wilson operator extended along time direction,
\begin{equation}
P e^{i \int_{x^0}^{x^0+L^0}dx'^0 a_0}.
\label{10.42}
\end{equation}
According to \eqref{10.12}, we see that under  \eqref{10.41}, this object transforms as
\begin{eqnarray}
P e^{i \int_{x^0}^{x^0+L^0}dx'^0 a_0} ~&\rightarrow&~ \tilde{U}(x^0+L^0)  P e^{i \int_{x^0}^{x^0+L^0}dx'^0 a_0} \tilde{U}^{\dagger}(x^0)\nonumber\\
~&\rightarrow&~ e^{\frac{2\pi i k}{N}}\tilde{U}(x^0)  P e^{i \int_{x^0}^{x^0+L^0}dx'^0 a_0} \tilde{U}^{\dagger}(x^0),
\label{10.43} 
\end{eqnarray}
where we have used \eqref{10.40}. Then, taking the trace leads to 
\begin{equation}
\text{Tr}P e^{i \int_{x^0}^{x^0+L^0}dx'^0 a_0} ~\rightarrow~ e^{\frac{2\pi i k}{N}} \text{Tr}P e^{i \int_{x^0}^{x^0+L^0}dx'^0 a_0}.
\label{10.44}
\end{equation}
In other words, the Wilson line in the fundamental representation is charged under \eqref{10.41}, with the charge being an element of the center $\mathbb{Z}_N$.


\subsubsection{Adjoint Representation}

Let us construct now a Wilson line with the generators taken in the adjoint representation. To this, we can use the fact that the direct product of fundamental and anti-fundamental representations of $SU(N)$ can be decomposed into irreducible parts according to 
\begin{equation}
\underbrace{N}_{\text{dim}=N} \otimes \underbrace{\bar{N}}_{\text{dim}=N} = \underbrace{\openone}_{\text{dim}=1} \oplus \underbrace{A}_{\text{dim}=N^2-1}.
\label{10.45}
\end{equation}
This expression provides us a useful relation for the indices of the adjoint representation in terms of fundamental indices, known as double-index representation. To see this, we consider fields $\phi^i$ and $\phi_i$ transforming according to the fundamental and anti-fundamental representations, namely, $\phi^{'i}=U^i\,_j \phi^j$ and $\phi'_i=(U^*)_i\,^j \phi_j$. In terms of components, \eqref{10.45} reads
\begin{equation}
\phi^i \phi_j = \frac{1}{N}\delta^i_j \phi^2+\left(\phi^i \phi_j - \frac{1}{N}\delta^i_j \phi^2\right).
\label{10.46}
\end{equation}
The term inside the brackets corresponds to the tensor product that transforms in the adjoint representation. We define
\begin{equation}
\phi_{adj}^{\overline{ij}} \equiv \phi^i \phi_j - \frac{1}{N}\delta^i_j \phi^2.
\label{10.47}
\end{equation}
With this, we notice that fields belonging to the adjoint representation are invariant under the center symmetry, since their transformation involve the product of $U$ and $U^*$.

Under a generic $SU(N)$ transformation, it follows that
\begin{eqnarray}
\phi_{adj}^{'\overline{ij}} &=&  \phi^{'i} \phi'_j - \frac{1}{N}\delta^i_j \phi^{'2}\nonumber\\
&=& U^{i}\,_k (U^*)_j\,^{l} \phi^k \phi_l -\frac{1}{N}\delta^i_j \phi^{2}\nonumber\\
&=& U^{i}\,_k (U^*)_j\,^{l} \left( \phi^k \phi_l-\frac{1}{N}\delta^k_l \phi^2+\frac{1}{N}\delta^k_l \phi^2\right)-\frac{1}{N}\delta^i_j \delta_k^l  \left( \phi^k \phi_l-\frac{1}{N}\delta^k_l \phi^2+\frac{1}{N}\delta^k_l \phi^2\right)\nonumber\\
&=& \left(U^{i}\,_k (U^*)_j\,^{l}-\frac{1}{N}\delta^i_j \delta_k^l  \right) \left( \phi^k \phi_l-\frac{1}{N}\delta^k_l \phi^2\right),
\label{10.48}
\end{eqnarray}
where we have used that 
\begin{equation}
\frac{1}{N}U^{i}\,_k (U^*)_j\,^{l} \delta^k_l \phi^2-\frac{1}{N^2}\delta^i_j \delta_k^l \delta^k_l \phi^2=0.
\label{10.49}
\end{equation}
Relation \eqref{10.48} implies that any $SU(N)$ matrix in the adjoint representation can be written as
\begin{equation}
U^{\overline{ij}}_{\overline{kl}}\equiv U^{i}\,_k (U^*)_j\,^{l}-\frac{1}{N}\delta^i_j \delta_k^l,
\label{10.50}
\end{equation}
so that 
\begin{equation}
\phi_{adj}^{'\overline{ij}} =U^{\overline{ij}}_{\overline{kl}}\, \phi_{adj}^{\overline{kl}}.
\label{10.51}
\end{equation}

Let us use the above results to construct a Wilson line in the adjoint representation. In the fundamental representation, we can write
\begin{equation}
[P e^{i \oint a}]^i\,_j \in SU(N).
\label{10.52}
\end{equation}
According to \eqref{10.50}, it follows that the Wilson line in the adjoint representation is
\begin{equation}
[P e^{i \oint a}]^{\overline{ij}}_{\overline{kl}}=[P e^{i \oint a}]^i\,_k [(P e^{i \oint a})^*]_j\,^l -\frac{1}{N}\delta^i_j \delta_k^l.
\label{10.53}
\end{equation}
Finally, taking the trace amounts to setting $i=k$ and $j=l$, 
\begin{eqnarray}
\text{Tr}_{adj} \left(P e^{i \oint a} \right)&=& [P e^{i \oint a}]^i\,_i [(P e^{i \oint a})^*]_j\,^j -\frac{1}{N}\delta^i_j \delta_i^j\nonumber\\
&=& \left|  P e^{i \oint a}  \right|^2-1.
\label{10.54}
\end{eqnarray} 
Therefore, we see that the Wilson line in the adjoint representation is invariant under the  1-form center $\mathbb{Z}_N$ symmetry.


\subsection{'t Hooft Lines}

Now we study  't Hooft lines, which represent the trajectory of "magnetically" charged probe particles \cite{Goddard:1976qe} (see also  \cite{Tong:2018}). Perhaps the simplest way to introduce the 't Hooft lines in the non-Abelian case is to take a line extended along time direction, and then consider its effect on a $S^2$ surface that links with such line. As there is no electric flux, $f_{0i}=0$, and the magnetic flux is produced by a  magnetic field imitating the Abelian case,
\begin{equation}
b^{i}=\frac{1}{2}\epsilon^{ijk}f_{jk} = \frac{x^i}{4 \pi |\vec{x}|^3} Q(x) ~~~\Leftrightarrow~~~ f^{ij} =  \epsilon^{ijk} \frac{x^k}{4 \pi |\vec{x}|^3} Q(x).
\label{10.55}
\end{equation} 
The algebra-valued object $Q(x)=Q^a(x)T^a$ plays the role of the magnetic charge of the 't Hooft line. It must carry in principle dependence on coordinates since under gauge transformations $f_{ij}\rightarrow U f_{ij}U^{\dagger}$, and consequently $Q$ must transform accordingly, i.e.,  $Q\rightarrow U Q U^{\dagger}$. 

The equations of motion \eqref{10.6b} and \eqref{10.6e} reduce to
\begin{equation}
D_i f^{ij}=0~~~\text{and}~~~ \epsilon^{ijk} D_i f_{jk}=0,
\label{10.56}
\end{equation}
which, in turn, imply that 
\begin{equation}
D_i Q= \partial_i Q -i [a_i,Q] =0.
\label{10.57}
\end{equation}

Potentials producing the configuration in \eqref{10.55} can be constructed in a simple way in spherical coordinates. We recall that we need two charts to cover $S^2$. Considering first a chart covering the north pole, we can take the potential as
\begin{equation}
a_r^N=a_{\theta}^N=0~~~\text{and}~~~a_{\phi}^N=\frac{Q}{4\pi r } \frac{(1-\cos\theta)}{\sin\theta},
\label{10.58}
\end{equation}
with a {\it constant} matrix algebra-valued $Q$. This is consistent with \eqref{10.57}, namely, $D_{\phi} Q = \partial_{\phi} Q -i [a_{\phi},Q]=0$.
In terms of differential forms, this reads
\begin{equation}
a^N=\vec{a}^N \cdot d\vec{x}= a_r^N dr + r\, a_{\theta}^N d\theta + r \sin\theta\, a_{\phi}^N d\phi  =\frac{Q}{4\pi}(1-\cos\theta)d\phi. 
\label{10.59}
\end{equation}
Notice that we can write the magnetic field in \eqref{10.55} as
\begin{eqnarray}
b^i = \epsilon^{ijk}\partial_{j}a_k - i \epsilon^{ijk}a_j a_k.
\label{10.60}
\end{eqnarray}
Then we convert this to spherical coordinates. The term $ \epsilon^{ijk}a_j a_k$ vanishes for the fields given in \eqref{10.58} because only the component $a_{\phi}^N$ is nonvanishing. The contribution of the curl, 
\begin{equation}
\vec{b}=\vec{\nabla}\times \vec{a}^N= \hat{r} \frac{1}{r \sin\theta}\left[\frac{\partial}{\partial\theta}(\sin\theta\, a_{\phi}^N)-\frac{\partial a_{\theta}^N}{\partial\phi}\right]+\cdots,
\label{10.61}
\end{equation}
produces the magnetic field in \eqref{10.55}. The same reasoning applies to the chart covering the south pole. In this case, the potential reads
\begin{equation}
	a_r^S=a_{\theta}^S=0~~~\text{and}~~~a_{\phi}^S=-\frac{Q}{4\pi r } \frac{(1+\cos\theta)}{\sin\theta},
	\label{10.62}
\end{equation}
which also gives the magnetic field \eqref{10.55}. In the overlapping region ($\theta=\frac{\pi}{2}$) they can differ at most by a gauge transformation. 

With a suitable gauge transformation $Q\rightarrow U Q U^{\dagger}$ we can diagonalize $Q$, so that it can be written as a linear combination of the Cartan generators\footnote{A nice book on Lie algebras can be found in \cite{Georgi:1999wka}. In Appendix \ref{wr}, we review some useful properties.} since the diagonal form is still algebra-valued, 
\begin{equation}
Q = \vec{m} \cdot \vec{H},
\label{10.63} 
\end{equation}
where the coefficient $m_i$ are determined by consistency. To see this, we can proceed in analogy to the equations \eqref{5.9b} and \eqref{5.9c} of the Abelian case or, alternatively, we can consider a Wilson line along the overlapping region ($\theta=\frac{\pi}{2}$). In this region, a closed path should lead to the same result using either $a^N$ or $a^S$,
\begin{eqnarray}
 \text{Tr} P \exp \left(i \int_{0}^{2\pi}a_{\phi}^N \right) &=&  \text{Tr} P \exp \left(i \int_{0}^{2\pi}a_{\phi}^S \right)\nonumber\\
 \text{Tr} P e^{\frac{iQ}{2}}&=&\text{Tr} P e^{\frac{-iQ}{2}}\nonumber\\
  \text{Tr} e^{ \frac{i \vec{m}\cdot\vec{H}}{2}}&=& \text{Tr} e^{ \frac{-i \vec{m}\cdot\vec{H}}{2}},
  \label{10.64}
\end{eqnarray}
where in the last step we have discarded the path ordering since all the Cartan generators are simultaneously commuting. Taking the trace in the basis that diagonalizes the Cartan generators, this leads to a consistency condition in terms of the weights $\vec{\mu}$ of the corresponding representation,
\begin{equation}
e^{i \vec{m}\cdot \vec{\mu}}=1~~~\Rightarrow~~~ \vec{m}\cdot \vec{\mu} = 2\pi \mathbb{Z}.
\label{10.65}
\end{equation}
At this point is useful to recall the relation between weights and roots of the $SU(N)$ algebra, 
\begin{equation}
	2 \frac{\vec{\alpha}\cdot \vec{\mu}}{\vec{\alpha}^2} \in \mathbb{Z}.
	\label{10.66}
\end{equation}
For convenience we review the derivation of this relation in appendix \ref{wr}.
Therefore, condition \eqref{10.65} is satisfied for any representation labeled by $\vec{\mu}$  provided that 
\begin{equation}
\vec{m} \equiv 2\pi \frac{2\vec{\alpha}}{\vec{\alpha}^2},
\label{10.67} 
\end{equation}
i.e., the magnetic charges are associated with the roots. In other words, the t' Hoot lines must be taken in the adjoint representation. Summarizing, while the Wilson lines can be taken in any representation, the 't Hooft lines must be taken in the adjoint one. This leads to an imbalance of the number of electric and magnetic line operators.


\subsection{$SU(N)$ vs $SU(N)/\mathbb{Z}_N$}

With the previous understanding of Wilson and 't Hooft operators, we can address the important question about the difference between $SU(N)$ and $SU(N)/\mathbb{Z}_N$ gauge theories. From the point of view of the (local) action, both theories are the same. However, the spectrum of line operators is different. To highlight the differences, it is convenient to proceed similarly to the Abelian case and parametrize the field strength and its Hodge dual as
\begin{equation}
f= da -i a \wedge a ~~~\text{and}~~~ *f = d\tilde{a} -i \tilde{a} \wedge \tilde{a}.
\label{10.68}
\end{equation}
In terms of the fields $a$ and $\tilde{a}$, we construct the Wilson and t'Hooft operators as
\begin{equation}
W[\mathcal{C}] = \text{Tr}_{R_e} P e^{i \oint a}~~~\text{and}~~~T[\mathcal{C}] = \text{Tr}_{R_m} P e^{i \oint \tilde{a}},
\label{10.69}
\end{equation}
with $R_e$ and $R_m$ corresponding to the representations in which they are taken. In this way, we may have in general electric and magnetic $\mathbb{Z}_N$ 1-form symmetries, 
\begin{equation}
(\mathbb{Z}_N)_e^{(1)}~~~\text{and}~~~ (\mathbb{Z}_N)_m^{(1)}.
\label{10.70}
\end{equation}

Now we specify the gauge group. In the $SU(N)$ case, as we have discussed above, the Wilson lines can be taken in any representation $R_e$ labeled by the weights $\vec{\mu}$,  whereas the magnetic charges are labeled by the roots $\vec{m}$, meaning that the t' Hooft operators must be taken in the adjoint representation. 
In the case of $SU(N)/\mathbb{Z}_N$ gauge group, any object that is charged under the center symmetry is not allowed, restricting so the possible Wilson lines. In fact, only Wilson lines in the adjoint representation (more generally, tensor products of the adjoint representation) are acceptable, which means that they are labeled in this case by the roots. Consequently, the consistency condition \eqref{10.65} implies that the 't Hooft lines are labeled by the weights, so that they can be taken in any representation $R_m$. The role of the weights and roots is exchanged in $SU(N)$ and $SU(N)/\mathbb{Z}_N$ groups. 

We can summarize the above picture in terms of the 1-form symmetries in each case,
\begin{equation}
SU(N):~(\mathbb{Z}_N)_e^{(1)}  ~~~\text{and}~~~ SU(N)/\mathbb{Z}_N:~(\mathbb{Z}_N)_m^{(1)}.
\label{10.71}
\end{equation}


\section{Spontaneous Breaking of Higher-Form Symmetries}\label{Sec10}

The last topic of these notes is the spontaneous symmetry breaking of higher-form symmetries \cite{Gaiotto:2014kfa,Lake:2018dqm,Hofman:2018lfz}. For concreteness, we shall consider the case of 1-form symmetries in gauge theories, whose charged objects are Wilson and 't Hooft lines. 

The key point is to identify the vacuum expectation value of the Wilson loop as the order parameter for the electric 1-form symmetry, which distinguishes different phases.
Given a closed curve $\mathcal{C}$, the expectation value $\langle W[\mathcal{C}]\rangle$ typically depends on geometric properties like the area enclosed by $\mathcal{C}$ or its perimeter, 
\begin{equation}
\langle W[\mathcal{C}] \rangle \sim e^{- \text{Area}[\mathcal{C}]}~~~\text{or}~~~\langle W[\mathcal{C}] \rangle \sim e^{- \text{Perimeter}[\mathcal{C}]}.
\label{11.1}
\end{equation}
Distinct decaying behaviors naturally signal different phases. For a large loop $\mathcal{C}$, the area law decays much faster than the perimeter law, so that effectively we have
\begin{equation}
\langle W[\mathcal{C}] \rangle \sim e^{- \text{Area}[\mathcal{C}]} \rightarrow 0, 
\label{11.2}
\end{equation}
while
\begin{equation}
\langle W[\mathcal{C}] \rangle \sim e^{- \text{Perimeter}[\mathcal{C}]} \neq 0.
\label{11.3}
\end{equation}
Another typical dependence, with a decay even weaker than the perimeter law is the so-called Coulomb behavior, which is a scale-invariant dependence on the parameters of the loop. We shall consider this explicitly a little later. 

In analogy with the case of ordinary symmetries, the perimeter and Coulomb laws are interpreted as implying a nonvanishing value of the order parameter and then are associated with phases where the 1-form symmetry is spontaneously broken.  As we shall see, when the 1-form symmetry is continuous, this leads to the existence of Goldstone excitations (the photon!). 
On the other hand, the area law is interpreted as meaning a vanishing value of the order parameter, corresponding to a symmetric phase. 
This perspective is quite illuminating and, in particular, allows to reinterpret the problem of confinement in terms of spontaneous breaking of a 1-form symmetries. 
\begin{figure}
	\centering
	\includegraphics[scale=.6,angle=90]{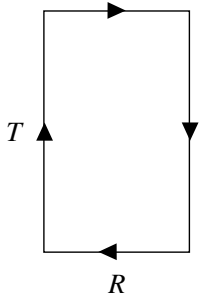}
	\caption{Wilson line describing a pair of static particles.}
	\label{w}
\end{figure}

To appreciate this, let us start by discussing the relation of the expectation value of the Wilson loop in Euclidean spacetime with the static potential between two charged probe particles. The idea is to consider a loop as shown in Fig. \ref{w}, which has a simple physical interpretation: a pair of opposite charges is created in the remote past by a source which is adiabatically turned on, and then they are slowly separated apart from each other at a distance $R$. After a long time $T$ the pair is annihilated, again adiabatically.  

We are interested in the corresponding vacuum expectation value
\begin{equation}
\langle  W[\mathcal{C}] \rangle=  \langle\text{Tr}P e^{i \oint_{\mathcal{C}}a} \rangle,
\label{11.4}
\end{equation}
computed in the Euclidean spacetime. We will follow here \cite{Makeenko:2009dw}.
As this object is gauge invariant, we can choose a convenient gauge. Let us pick up the axial gauge, $a_0=0$, so that there is no contribution of the pieces of $\mathcal{C}$ along the time direction $T$. Also, without loss of generality, we consider that $R$ is along direction 1. In this situation, the above expression reduces to
\begin{equation}
\langle  W[\mathcal{C}] \rangle= \langle [P  e^{i \int_0^R dx^1 a_1(T,x^1,\ldots)}]^i\,_j [P  e^{ i \int_R^0 dx^1 a_1(0,x^1,\ldots)}]^j\,_i  \rangle.
\label{11.5}
\end{equation}
To simplify the notation, it is convenient to define
\begin{equation}
\psi^i\,_j (T) \equiv  [P  e^{i \int_0^R dx^1 a_1(T,x^1,\ldots)}]^i\,_j.
\label{11.6}
\end{equation}
With this, \eqref{11.5} becomes
\begin{equation}
\langle  W[\mathcal{C}] \rangle= \langle \psi^i\,_j (T) \psi^{\dagger j}\,_i(0) \rangle. 
\label{11.7}
\end{equation}
Recalling that the Euclidean time $T$ is related to the real time according to $t\rightarrow - i T$, the time evolution is $\psi^i\,_j (T)=e^{H T} \psi^i\,_j (0)e^{-H T}$. With this, and inserting a complete set of energy eigenstates $\ket{n}$ in \eqref{11.7}, it follows that
\begin{eqnarray}
\langle  W[\mathcal{C}] \rangle &=& \sum_n e^{-T E_n(R)} \langle \psi^i\,_j (0) \ket{n}\bra{n} \psi^{\dagger j}\,_i(0) \rangle\nonumber\\
&=&  \sum_n e^{-T E_n(R)} | \langle \psi^i\,_j (0) \ket{n}|^2.
\label{11.8}
\end{eqnarray}
In the limit $T\rightarrow\infty$, only the lowest-energy state provides a significant contribution,
\begin{equation}
\langle  W[\mathcal{C}] \rangle \sim e^{-T E_0(R)}. 
\label{11.9}
\end{equation}
As the charges are static, the energy reduces to the potential, and we finally find
\begin{equation}
V(R)= - \lim_{T\rightarrow \infty} \frac{1}{T} \ln \langle  W[\mathcal{C}] \rangle. 
\label{11.10}
\end{equation}
So the behavior of the loop Wilson dictates the form of the static potential between charges.

\subsubsection{Phases of Gauges Theories}

We discuss now the typical behaviors of the expectation value of the Wilson loop associated with Fig. \ref{w}. Suppose it behaves as the area law,  
\begin{equation}
\langle  W[\mathcal{C}] \rangle \sim e^{-\sigma  T R},
\label{11.11}
\end{equation}
where $\sigma$ is a dimensionful constant. According to \eqref{11.10}, this leads to a linear potential,
\begin{equation}
V(R)=\sigma R.
\label{11.12}
\end{equation}
Therefore, the energy required to separate charges grows linearly with the distance $R$, so that the charges turn out to be confined. In other words, the area law for $\langle W[\mathcal{C}]\rangle$ is an indicative of {\it confinement}.

Next we consider the perimeter law,
\begin{equation}
\langle  W[\mathcal{C}] \rangle \sim e^{-\rho (T+ R)},
\label{11.13}
\end{equation}
where $\rho$ is a dimensionful constant. This behavior leads to a  constant potential  
\begin{equation}
	V(R)=\rho.
	\label{11.14}
\end{equation}
Since the energetic cost to separate charges at large distance is finite, this potential does not confine charges. We rephrase this by saying that the perimeter law corresponds to a {\it deconfining} phase.

Finally, we consider the Coulomb or scale-invariant law. In this case, $\langle W[\mathcal{C}]\rangle$ decays slower than the perimeter law, depending on the dimensionless ratios $T/R$ or $R/T$, 
\begin{equation}
\langle W[\mathcal{C}]\rangle \sim e^{- \alpha \frac{T}{R} -\beta \frac{R}{T}},
\label{11.15}
\end{equation}
where $\alpha$ and $\beta$ are dimensionless constants. The corresponding potential is
\begin{equation}
V(R)= \frac{\alpha}{R},
\label{11.16}
\end{equation}
which is precisely the Coulomb potential. Naturally, this also corresponds to a {\it deconfining} phase.


\subsection{Continuous 1-Form Symmetry and Goldstone Excitations}

To find the Goldstone excitations, we can follow precisely the same reasoning we did in the case of ordinary symmetries in Sec. \eqref{ssb0}.
The starting point is the Ward identity \eqref{3.20}, which we rewrite here with $\langle \delta W[\mathcal{C}]\rangle=-i q_e \langle W[\mathcal{C}]\rangle$, 
\begin{equation}
\langle \partial_{\mu}J^{\mu\nu}(x) \,W[\mathcal{C}] \rangle = - q_e \int_{\mathcal{C}} dy^{\nu} \delta^{(D)}(x-y) \langle W[\mathcal{C}]\rangle.
\label{11.17}
\end{equation}
Taking the Fourier transform$\int d^Dx e^{ipx}$, this leads to
\begin{equation}
i p_{\mu}\langle J^{\mu\nu}(p) W[\mathcal{C}]\rangle = q_e f^{\nu}(p;\mathcal{C}) \langle W[\mathcal{C}]\rangle,
\label{11.18}
\end{equation}
where we have defined 
\begin{equation}
f^{\nu}(p;\mathcal{C})\equiv \int_{\mathcal{C}} dy^{\nu} e^{ipy}.
\label{11.19} 
\end{equation}
This object has some important properties. First, it is in general nonvanishing at $p=0$, i.e., 
\begin{equation}
f^{\nu}(0;\mathcal{C})\equiv \int_{\mathcal{C}} dy^{\nu}\neq 0.
\label{11.20}
\end{equation}
Second, it satifies
\begin{eqnarray}
p_{\nu}f^{\nu}(p;\mathcal{C})&=& \int_{\mathcal{C}} dy^{\nu} p_{\nu} e^{ipy}\nonumber\\
&=& -i \int_{\mathcal{C}} dy^{\nu} \partial_{\nu}  e^{ipy} =0,
\label{11.21}
\end{eqnarray}
for any closed curve $\mathcal{C}$.

With this, we consider \eqref{11.18} in the limit $p\rightarrow 0$, 
\begin{equation}
\lim_{p\rightarrow 0} i p_{\mu}\langle J^{\mu\nu}(p) W[\mathcal{C}]\rangle = q_e f^{\nu}(0;\mathcal{C}) \langle W[\mathcal{C}]\rangle.
\label{11.22}
\end{equation}
Whenever $\langle W[\mathcal{C}]\rangle \neq 0$, as it happens in the case of Coulomb and perimeter laws, the correlation function $\langle J^{\mu\nu}(p) W[\mathcal{C}]\rangle $ must have a pole at $p=0$,
\begin{equation}
\langle J^{\mu\nu}(p) W[\mathcal{C}]\rangle \sim \frac{p^{\mu}f^{\nu}(p,\mathcal{C})-p^{\nu}f^{\mu}(p,\mathcal{C})}{p^2}.
\label{11.23}
\end{equation} 
This, in turn, implies that there are massless excitations in the spectrum. These excitations are the Goldstone bosons following from the spontaneous breaking of the 1-form symmetry.


\subsection{QED in $D=4$}

The expectation value of the Wilson loop in four-dimensional free QED behaves according to the Coulomb law, leading to the deconfinement of charges. The above discussion then implies that there are Goldstone excitations in the spectrum. We can understand in a very simple way that the photons are precisely the Goldstone excitations coming from the spontaneous breaking the 1-form symmetry. In fact, we know that the corresponding conserved current $J^{\mu\nu}$ creates Goldstone excitations from the vacuum in the broken phase\footnote{A very nice exposition on spontaneous symmetry breaking can be found in \cite{Beekman:2019pmi}.}, 
\begin{equation}
\ket{\text{Goldstone}} \sim J^{\mu\nu}(x) \ket{0}.
\label{11.24}
\end{equation}
We recall that $J^{\mu\nu}=f^{\mu\nu}$. To proceed we need to enter a little bit into the canonical quantization structure. For the theory covariantly quantized in the Feynman gauge, i.e., we consider the Lagrangian with the addition of the gauge-fixing term $-\frac{\xi}{2}(\partial_{\mu} a^{\mu})^2$ and set $\xi=1$, the free field expansion satisfying $\partial^2 a^{\mu}=0$ is
\begin{equation}
a^{\mu}(x)=\frac{1}{(2\pi)^{\frac32}} \int \frac{d^3 \vec{p}}{2 |\vec{p}|} \sum_{\lambda=1}^4 e_{\lambda}^{\mu}(p)\left[a_{\lambda}(p)e^{-ipx} + a_{\lambda}^{\dagger}(p)e^{ipx} \right],
\label{11.25}
\end{equation}
where $e_{\lambda}^{\mu}(p)$ are four linearly independent polarization vectors. Not all polarizations are physical since some of them produce states that do not satisfies the selection rule $\partial_{\mu}a^{\mu}\ket{\text{phy}}=0$, or correspond simply to gauge degrees of freedom (zero norm states). Let us say that the physical polarizations are $\lambda=1,2$. Then, a single photon state is created by 
\begin{equation}
\ket{\lambda, \vec{p}} = a_{\lambda}^{\dagger}(p)\ket{0},~~~\lambda=1,2,
\label{11.26}
\end{equation}
with the creation and annihilation operators satisfying 
\begin{equation}
[a_{\lambda}(p),a_{\lambda'}^{\dagger}(p')]= 2 |\vec{p}| \delta_{\lambda,\lambda'} \delta^{(3)}(\vec{p}-\vec{p}'), ~~~\lambda, \lambda'=1,2.
\label{11.27}
\end{equation}
Now we can compute the matrix element between $\ket{\lambda, \vec{p}}$ and $\ket{\text{Goldstone}}\sim f^{\mu\nu}(x) \ket{0}$, 
\begin{equation}
\bra{0}f^{\mu\nu}(x) \ket{\lambda, \vec{p}}= \frac{i}{(2\pi)^{\frac32}} \left[e^{\mu}_{\lambda}(p) p^{\nu} -e^{\nu}_{\lambda}(p) p^{\mu}\right]e^{-ipx} \neq 0,
\label{11.28}
\end{equation}
which shows that the Goldstone excitation has nonvanishing overlap with (and only with) a single photon state. Therefore, this implies that the Goldstone excitation is the photon itself.


\section{Final Remarks}\label{Sec11}

As we have tasted along these notes, higher-form symmetries pervade gauge theories and thus figure as a fundamental ingredient in the modern perspective of effective field theories, which are largely grounded on the gauge structure. In a broader context, the subject of generalized symmetries (encompassing higher-forms and the other forms of symmetries) possesses some features that place it as one of the cornerstones of modern physics. 

Generalized symmetries have been a source of new advances and new results in several directions. New forms of symmetries naturally lead to more constraints in the underlying theories.
At the same time, they provide a deeper understanding of many known results and also shed light on some hard problems. For example, as we have discussed, they enable us to understand the photon as a Goldstone excitation and lead to a reformulation of confinement problem in terms of spontaneous breaking of a higher-form symmetry. 

In addition, generalized symetries are a meeting point of different areas of physics, placing them under a unified perspective. A remarkable example is fracton physics, which brings together aspects of quantum field theory, quantum computing, and topological phases of matter, whose exotic patterns of higher-form symmetries challenge the construction of effective field theories. Nevertheless, much progress has been done in this direction (see for example \cite{Slagle:2017wrc,Bulmash:2018lid,Pretko:2018jbi,Gromov:2018nbv,You:2019ciz,Seiberg:2019vrp,Slagle:2020ugk,Seiberg:2020bhn,Seiberg:2020cxy,Fontana:2020tby}).
Still in the context of topological phases, generalized symmetries lead to an enlargement of the Landau paradigm (based on symmetry breaking) to encompass topological order, which can then be interpreted as spontaneous breaking of higher-form symmetries  \cite{Wen:2018zux,Iqbal:2021rkn}.

Naturally, that type of interplay is a hallmark of deep ideas in physics, reflecting its universality across the fields. The result is a very fruitful cross-fertilization with far-reaching consequences.


\section{Acknowledgments}

I would like to thank Cameron Gibson and Carlos Hernaski for carefully reading the manuscript, pointing out several typos, and also for many helpful comments. This work is partially supported by CNPq.

		
\appendix

\section{Differential Forms}\label{diff}

In a $D$-dimensional manifold, a $p$-form ($p\leq D$) is expressed as
\begin{equation}
	\Omega_p = \frac{1}{p!}\omega_{\mu_1\mu_2\ldots\mu_p} dx^{\mu_1}\wedge\cdots\wedge dx^{\mu_p}.
	\label{2.1}
\end{equation} 
Let us recall some basic operations with differential forms \cite{bertlmann2000anomalies,nakahara2018geometry}. The exterior derivative is defined as
\begin{equation}
	d\Omega\equiv \frac{1}{p!}\partial_{\alpha}\omega_{\mu_1\mu_2\ldots\mu_p} dx^{\alpha}\wedge dx^{\mu_1}\wedge\cdots\wedge dx^{\mu_p}.
	\label{2.2}
\end{equation}
The Hodge dual operation $*$ is defined through
\begin{equation}
	*\Omega_p\equiv  \frac{1}{p!}\omega_{\mu_1\mu_2\ldots\mu_p} *dx^{\mu_1}\wedge\cdots\wedge dx^{\mu_p},
	\label{2.3}
\end{equation}
with the dual of the antisymmetrized product being
\begin{equation}
	*dx^{\mu_1}\wedge\cdots\wedge dx^{\mu_p}\equiv \frac{1}{(D-p)!}\epsilon^{\mu_1\ldots\mu_p}{}_{\mu_{p+1}\ldots\mu_D} dx^{\mu_{p+1}}\wedge\cdots\wedge dx^{\mu_D}.
	\label{2.4}
\end{equation}
Plugging this back in \eqref{2.3}, we find
\begin{equation}
	*\Omega=\frac{1}{p!} \frac{1}{(D-p)!}  \omega_{\mu_1\mu_2\ldots\mu_p} \epsilon^{\mu_1\ldots\mu_p}{}_{\mu_{p+1}\ldots\mu_D} dx^{\mu_{p+1}}\wedge\cdots\wedge dx^{\mu_D}.
	\label{2.5}
\end{equation}
Finally, we recall the Stokes theorem, which establishes the relation
\begin{equation}
	\int_{\mathcal{M}}d\omega = \int_{\partial\mathcal{M}} \omega,
	\label{2.5a}
\end{equation}
where $\mathcal{M}$ is a manifold with boundary $\partial\mathcal{M}$. The Stokes theorem encompasses the usual theorems of calculus. For example, in $\mathbf{R}^3$, if we pick up a 1-form $\omega=\omega_{\mu}dx^{\mu}$, the Stokes theorem gives
\begin{equation}
	\int_S \nabla\times \vec{\omega} \cdot d\vec{S} = \oint_C \vec{\omega}\cdot d\vec{l}~~~(\text{Stokes' theorem}), 
\end{equation}
where the curve $C$ is the boundary of the surface $S$. For $\omega=\frac{1}{2}\omega_{\mu\nu}dx^{\mu}\wedge dx^{\nu}$, 
\begin{equation}
	\int_V  \nabla\cdot \vec{\omega}\, dV =\oint_{S}\vec{\omega}\cdot d\vec{S}~~~(\text{Gauss' theorem}),
\end{equation}
where $\omega^{\mu}=\epsilon^{\mu\nu\rho}\omega_{\nu\rho}$ and the surface $S$ is the boundary of the volume $V$.


\section{Quick Review of Weights and Roots}\label{wr}

The analysis of line operators in the non-Abelian case requires going through a little further on the Lie algebra structure and representations \cite{Georgi:1999wka,DiFrancesco:1997nk}. We need to consider the so-called weights of the representation. They are the eigenvalues of the so-called Cartan subalgebra, which is the maximum set of self-commuting generators $T^a$, denoted by $H^i$, 
\begin{equation}
	[H^i,H^j]=0,~~~i=1,\ldots, r,
	\label{10.15}
\end{equation}
where $r$ is the rank of the algebra. They can be simultaneously diagonalized, 
\begin{equation}
	H^i \ket{\mu,R}= \mu_i \ket{\mu,R},
	\label{10.16}
\end{equation}
where the eigenvalues are called weights of the representation. All such weights span a lattice, referred to as the lattice weight $\Lambda_{w}(\mathfrak{g})$.

Special representations are the {\it fundamental} and the {\it adjoint}. The fundamental representation is $N$-dimensional. In general, matter fields are in this representation.
In the adjoint representation, the generators are 
\begin{equation}
	[T^a]_{bc} = -i \mathrm{f}_{abc}.
	\label{10.17}
\end{equation}
This is a $(N^2-1)$-dimensional representation. Acting on a state in the adjoint representation $T^a \ket{\psi, adj}$ amounts to
\begin{equation}
	\sum_c[T^a]_{bc} \ket{\psi, adj}_c,
	\label{10.18}
\end{equation}
namely, the states are specified by generator indices. This implies that we can associate generators with states,
\begin{equation}
	T^a ~~~\Leftrightarrow~~~ \ket{T^a},
	\label{10.19}
\end{equation}
with components $\ket{T^a}_b=\delta_{ab}$. Naturally, the scalar product is expected to be $\langle T^a | T^b\rangle \sim \text{Tr}(T^aT^b)$.

Linear combinations of the generators correspond to linear combination of states
\begin{equation}
	\alpha T^a + \beta T^b ~~~\Leftrightarrow~~~ \ket{\alpha T^a + \beta T^b}.
	\label{10.20}
\end{equation}
With this, we see that
\begin{eqnarray}
	T^a \ket{T^b}&=& \sum_c \ket{T^c}\bra{T^c} T^a \ket{T^b} \nonumber\\
	&=& \sum_c   [T^a]_{cb}  \ket{T^c}\nonumber\\
	&=& -i \mathrm{f}_{acb}  \ket{T^c}~~~(\text{sum convention})\nonumber\\
	&=& \ket{i \mathrm{f}_{abc} T^c}\nonumber\\
	&=& \ket{[T^a,T^b]}.
	\label{10.21}
\end{eqnarray}

Restricting to the Cartan subalgebra, this leads
\begin{equation}
	H^i \ket{H^j}=\ket{[H^i,H^j]}=0.
	\label{10.22}
\end{equation}
As all the Cartan generators can be simultaneously diagonalized, we can choose specific (in general, non-Hermitian) linear combinations of the generators outside the Cartan subalgebra as $E^{\alpha}$, so that the corresponding states satisfy
\begin{equation}
	H^i \ket{E^{\alpha}} = \alpha_i \ket{E^{\alpha}},
	\label{10.23}
\end{equation}
with the scalar product
\begin{equation}
	\langle E^{\alpha}| E^{\beta}\rangle= \lambda \text{Tr}(E^{\alpha\dagger} E^{\beta}) = \delta_{\alpha\beta} = \prod_i \delta_{\alpha_i \beta_i},
\end{equation}
where the constant $\lambda$ is chosen to ensure the normalization of the scalar product. 
The weights $\alpha_i$ of the adjoint representation are called {\it roots}, and also span a lattice, $\Lambda_{\text{root}}(\mathfrak{g}) \subset \Lambda_{w}(\mathfrak{g})$.
According to \eqref{10.21}, this means that
\begin{equation}
	[H^i,E^{\alpha}]= \alpha_i E^{\alpha}.
	\label{10.24}	
\end{equation}
Taking the Hermitian conjugate of \eqref{10.24}, it follows, 
\begin{equation}
	[H^i,E^{\alpha\dagger}]= - \alpha_i E^{\alpha\dagger},
	\label{10.25}	
\end{equation}
so that we can take
\begin{equation}
	E^{\alpha\dagger} = E^{-\alpha}.
	\label{10.26}
\end{equation}
All roots come in pairs $\pm \alpha_i$.

This is the analogue of the algebra of creation and annihilation operators of angular momentum algebra.  The pair of operators $E^{\alpha}$ and $E^{\alpha\dagger}$ plays the role of creation and annihilation operators, i.e., $H^i \leftrightarrow J_z$ and $E^{\alpha} \leftrightarrow J^{+},J^{-}$. Relation \eqref{10.25} yields to
\begin{equation}
	H^i E^{\pm\alpha} \ket{\mu,R}= (\mu_i\pm \alpha_i) E^{\pm\alpha} \ket{\mu,R}.
	\label{10.27}
\end{equation}
This equation is true for any representation $R$, but it is particularly important for the adjoint representation. In fact, it implies that the state  $E^{\alpha}\ket{E^{-\alpha}}$ has vanishing weight, and thus it must be a linear combination of the Cartan generators, 
\begin{equation}
	E^{\alpha}\ket{E^{-\alpha}} = \ket{c_i H^i}.
	\label{10.28}
\end{equation}
The constants $c_i$ can be determined by taking the scalar product with $\ket{H^j}$ in both sides,
\begin{eqnarray}
	c_j &=& \bra{H^j} E^{\alpha} \ket{E^{-\alpha}}\nonumber\\
	&=&  \langle H^j | [E^{\alpha},E^{-\alpha}]\rangle \nonumber\\
	&=& \lambda \text{Tr} ( H^j [E^{\alpha},E^{-\alpha}] )\nonumber\\
	&=&  \lambda \text{Tr} ( E^{-\alpha} [H^j,E^{\alpha}] )\nonumber\\
	&=& \lambda \alpha^j \text{Tr} ( E^{-\alpha} E^{\alpha}) = \alpha_j.
	\label{10.29}
\end{eqnarray}
Therefore, this implies that
\begin{equation}
	[E^{\alpha}, E^{-\alpha}] = \alpha_i H^i. 
	\label{10.30}
\end{equation}

With this, we see that the operators 
\begin{equation}
	E^{\pm}\equiv \frac{1}{|\vec{\alpha}|} E^{\pm \alpha}~~~\text{and}~~~ E^3 \equiv \frac{1}{\vec{\alpha}^2}\, \vec{\alpha}\cdot \vec{H}
	\label{10.31}
\end{equation}
form a $SU(2)$ subalgebra. For any given representation $R$, this has an important consequence
\begin{equation}
	E^3 \ket{\mu,R}= \frac{\vec{\alpha}\cdot \vec{\mu}}{\vec{\alpha}^2}  \ket{\mu,R},
	\label{10.32}
\end{equation}
which implies that 
\begin{equation}
	2 \frac{\vec{\alpha}\cdot \vec{\mu}}{\vec{\alpha}^2} \in \mathbb{Z}.
	\label{10.33}
\end{equation}

		
\bibliography{refs_symmetries}

\end{document}